\documentclass{aa} % for a printer version
\usepackage{aalongtable}    
\usepackage{supertabular}    
\usepackage{graphicx}
\usepackage{epsfig}
\usepackage{rotating}
\usepackage{lscape}
\usepackage[authoryear]{natbib}
\bibpunct{(}{)}{;}{a}{}{,} % to follow the A&A style

\begin{document}

\title{Evidence for early disk-locking among low-mass members of the Orion Nebula Cluster\thanks{Based 
on the {\sc flames} Science Verification proposal 60.A-9145(A) and the {\sc flames} proposal 76.C-0524(A).} }
   
\author{K. Biazzo \inst{1,2,3} \and C. H. F. Melo \inst{2} \and L. Pasquini\inst{2} \and S. Randich\inst{1} 
\and J. Bouvier\inst{4} \and X. Delfosse\inst{4} }
\offprints{K. Biazzo}
\mail{kbiazzo@arcetri.astro.it}

\institute{INAF - Osservatorio Astrofisico di Arcetri, Largo E. Fermi 5, 50125 Firenze, Italy 
  \and ESO - European Southern Observatory, Karl-Schwarzschild-Str. 3, 85748 Garching bei M\"unchen, Germany
  \and INAF - Osservatorio Astrofisico di Catania, via S. Sofia 78, 95123 Catania, Italy
  \and Laboratoire d'Astrophysique, Observatoire de Grenoble, BP 53, 38041 Grenoble C\'edex 9, France
}

\date{Received / accepted }

\abstract
% Context:
{We present new high-resolution spectroscopic 
observations for 91 pre-main sequence stars in the Orion Nebular Cluster (ONC) with masses in the range $0.10-0.25M_\odot$ 
carried out with the multi-fiber spectrograph {{\sc flames}} attached to the UT2 at the Paranal Observatory.}
% Aims:
{Our aim is to better understand the disk-locking scenario in very low-mass stars.}
% Methods:
{We have derived radial velocities, projected rotational velocities, 
and full width at 10\% of the H$\alpha$ emission peak. Using published measurements of infrared excess ($\Delta(I_{\rm C}-K)$), 
as disk tracer and equivalent width of the nead-infrared \ion{Ca}{ii} line $\lambda$8542, mid-infrared difference 
[3.6]$-$[8.0] $\mu$m derived by {{\it Spitzer}} data, and 10\% H$\alpha$ width as diagnostic of the level of 
accretion, we have looked for any correlation between projected angular rotational velocity divided by the radius 
($v\sin i/R$) and presence of disk and accretion.} 
% Results:
{For 4 low-mass stars, the cross-correlation function is clearly double-lined indicating that the stars are SB2 systems. The 
distribution of rotation periods derived from our $v\sin i$ measurements is unimodal with a peak of few days, in agreement 
with previous results for $M<0.25M_\odot$. The photometric periods were combined with our $v\sin i$ to derive the equatorial 
velocity and the distribution of rotational axes. Our $<\sin i>$ is lower than the one expected for a random distribution, as 
previously found. We find no evidence for a population of fast rotators close to the break-up velocity. A clear correlation 
between $v\sin i/R$ and $\Delta(I_{\rm C}-K)$ has been found. While for stars with no circumstellar disk 
($\Delta(I_{\rm C}-K)<0.3$) a spread in the rotation rates is seen, stars with a circumstellar disk ($\Delta(I_{\rm C}-K)>0.3$) 
show an abrupt drop in their rotation rates by a factor of $\sim 5$. On the other hand, only a partial correlation between 
$v\sin i$ and accretion is observed when other indicators are used. The X-ray coronal activity level ($\log L_{\rm X}/L_{\rm bol}$) 
shows no dependence on $v\sin i/R$ suggesting that all stars are in a saturated regime limit. The critical velocity is probably 
below our $v\sin i$ detection limit of 9 km s$^{-1}$.}
% Conclusions: 
{The ONC low-mass stars in our sample, close to the hydrogen burning limit, at present seem to be not locked, but the 
clear correlation we find between rotation and infrared color excess suggests that they were locked once. In 
addition, the percentage of accretors seems to scale inversely to the stellar mass.}
   
	   
\titlerunning{Evidence for early disk-locking in ONC low-mass stars}
\authorrunning{K. Biazzo et al.}
\maketitle

\section{Introduction}
Surface rotation is a key observational parameter for stellar evolution because it is linked to the
internal angular momentum transport and to the mechanisms responsible for stellar angular momentum loss. 
Its initial conditions are set during the star formation process. 

There are four main ingredients describing the angular momentum (AM) evolution of $1 M_\odot$ star: {\it i) Solid 
body rotation.} Stars begin their lives as fully convective objects. Since the timescale for convective transport is 
much shorter than that for angular momentum loss, the stars probably rotate as solid bodies 
(\citealt{stassuntendrup2003}, and references therein). {\it ii) Disk-locking.} The idea of magnetic disk-locking 
originated with the theory developed by \cite{GhoshLamb1979} for neutron stars, but in the context of pre-main sequence 
(PMS) stars was proposed for the first time by \cite{Camenzind1990} and \cite{konig1991}. The stellar magnetic field 
threads the slowly rotating circumstellar disk of the young star, truncating it at several stellar radii 
(\citealt{bouvieretal2007}). Due to the difference in angular velocity, magnetic torques will transfer angular 
momentum from the star to the disk causing them to corotate, i.e. the spin-up torque on the star is exactly balanced 
by a spin-down torque transmitted by the field lines threading the disk beyond the corotation radius. In 
the end, the stellar rotational angular velocity equals the Keplerian angular velocity of the 
inner disk (\citealt{shu1994}). {\it iii) Stellar winds.} Angular momentum loss continues via a magnetized wind 
during the evolution to and on the main-sequence. The rate of angular momentum loss depends on whether the angular 
rotation rate is above the saturation limit (\citealt{barnessofia96}). {\it iv) Core-envelope decoupling.} Surface 
rotation is eventually replenished by AM brought from the rapid rotating core in a characteristic time $\tau_c$. 
How the decoupling characteristic depends on mass (and on other stellar parameters) is still unsettled. 

Although many details remain to be understood, these four ingredients seem to regulate the broad aspects of 
the AM evolution from PMS to solar age for $\sim 1 M_\odot$ stars. Among these four ingredients, the disk-locking 
hypothesis is believed to play a fundamental role in the early AM evolution during the T-Tauri phase, when young stars 
are expected to have circumstellar disks. This scenario has been supported through the years by different observational
photometric studies which show that T-Tauri stars with disks (CTTS) have significantly longer periods than
their disk-less counterparts (WTTS), i.e. they rotate at a small fraction (10--20\%) of the break-up velocity 
(\citealt{attridge1992, bouvieretal93, edwardsetal93, choiherbst96, marilli2007}). This indicates an efficient 
method for shedding AM, which is contrary to the expectation that these stars should spin close to break-up 
speed, having recently contracted from their natal clouds (\citealt{stassunetal99}).

In contrast to the $1 M_\odot$ mass regime, the AM evolution for lower mass stars is far from being 
understood. In particular, there are contrasting opinions regarding the distribution of rotational period 
($P_{\rm rot}$) for very low-mass stars ($M\la0.25 M_\odot$). From an observative point of view, \cite{stassunetal99} 
have challenged the disk-locking scenario in 254 stars of the young Orion OBIc/d association by showing that there 
is no evidence for the bimodal period distribution originally attributed to disk braking. In addition, no correlation 
between observed $P_{\rm rot}$ and accretion diagnostics and no differences between $P_{\rm rot}$ distributions of 
WTTS and CTTS were found, calling the standard disk-locking scenario into question. 
\cite{herbstetal00,herbstetal01,herbstetal02} obtain a bimodal period distribution 
for the Orion Nebula Cluster (ONC) confirming the findings of \cite{choiherbst96} and point out the 
dependence of the rotation on mass. They find a bimodal distribution of rotation periods for higher masses 
($M\ga0.25M_\odot$) with a gap near 4 days, while lower masses ($M\la0.25M_\odot$) have a unimodal distribution 
and generally spin faster. They show a statistically significant anti-correlation between IR excess emission 
and rotation: slower rotators are more likely to show evidence of circumstellar disks. \cite{rebull01} 
looked for the mass dependence in ONC flaking fields, but found only weak evidence for a 
change in rotational period distribution for low-mass stars. Later, \cite{rebulletal02} have shown that in stars 
of the ONC and NGC2264 AM is depleted very early during the T-Tauri phase and have concluded that disk-locking is 
indeed the most likely mechanism to be acting. \cite{hartmann02}, adopting accretion rates of 
$\dot M \sim 10^{-8.6} M_\odot$ yr$^{-1}$ taken from \cite{rebulletal00} for the ONC flaking fields for 
$M\sim0.3M_\odot$ and an initial rotation rate at 10\% of the break-up velocity, concluded that the time-scale 
needed to remove stellar angular momentum by disk-locking is of the order of the ONC age ($\sim 1-3$ Myr). In addition, 
he suggested that if these properties change with mass in a way to enhance the braking for higher masses, this could explain 
the bimodal distribution found in some studies. Then, \cite{Makidonetal2004} report for stars of NGC~2264 ($\sim 2-4$ Myr) 
no conclusive evidence that more slowly rotating stars show disk indicators or that faster rotating stars are less likely 
to show disk indicators. Later, \cite{Littlefairetal2005} find in IC~348 ($\sim 2-4$ Myr) $P_{\rm rot}$ bimodality at 
$M\ga0.25M_\odot$ and unimodality at $M\la0.25M_\odot$ and claim a strong mass effect as explanation. The same year, 
\cite{Lammetal2005} find evidences of disk-locking for stars of the same OB association, less pronounced for low 
masses ($M\la0.25M_\odot$). For this mass regime they speak about ``moderate angular momentum loss''. More recently, 
\cite{rebulletal2006} and \cite{CiezaBaliber2007} find that in Orion and NGC~2264 stars with long 
periods are more likely than those with short periods to have IR excesses suggestive of disks. Very recently, 
\cite{Nguyenetal2009} find in Taurus-Auriga and Chamaeleon I ($\sim 2$ Myr) that both accretors and non-accretors 
have similar distributions of $v \sin i$, while \cite{Rodri-Lede2009} confirm the result found by 
\cite{herbstetal00,herbstetal01,herbstetal02} in ONC. 

From a theoretical point of view, \cite{MattPudritz2004} critically examined the disk-locking theory 
and showed that in stars, such as CTTSs, where the magnetic field is strong, it is also highly twisted and disordered. 
The differential rotation between the star and disk naturally leads to an opening (i.e. disconnecting) of the 
magnetic field between the two. Consequently, the resulting spin-down torque on the star by the disk is significantly 
reduced, and the disk-locking model cannot account for accreting stars that spin slowly (e.g., about 10\% of the 
break-up velocity). Hence, the disk-locking scenario does not explain the angular momentum loss of the slow 
rotators. They conclude that, in order for accreting protostars to spin as slowly as 10\% of the break-up speed, there 
must be spin-down torques acting on the star other than those carried by magnetic field lines connecting the star to 
the disk. The presence of open stellar field lines leads them to the possibility that excess angular momentum is 
carried by a stellar wind along those open lines (\citealt{MattPudritz2004}). 

Therefore, it is clear that the angular momentum evolution for very low-mass stars is still not well understood. 
For a review concerning the rotation and angular momentum evolution of young and very low-mass stars from observative and 
theoretical points of view, see \cite{HerbstMundt2005}, \cite{Herbstetal2007}, and \cite{IrwinBouvier2009}.

In this work we measure H$\alpha$ excesses, projected rotational velocities ($v\sin i$), and radial velocities ($V_{\rm rad}$) 
of stars in the ONC with masses between 0.10 and 0.25 $M_\odot$. The ONC is the nearest large region of on-going star 
formation. Our study of this star-forming region is aimed at investigating whether there is any observational evidence supporting 
disk-locking in this still poorly studied mass range. More specifically, in order to shed more light on the disk-locking 
scenario for very low-mass stars, we look for any relation between rotation and accretion, and rotation and disk presence. 
We also look for the fast rotators population first reported by \cite{stassunetal99} 
and compare our rotational velocity distribution to the one derived by \cite{herbstetal00,herbstetal01,herbstetal02}.

\section{Observations, Data Reduction, and Analysis}
\subsection{Sample selection and Observations}

\begin{figure*}[th]
\center
       \includegraphics[width=12cm]{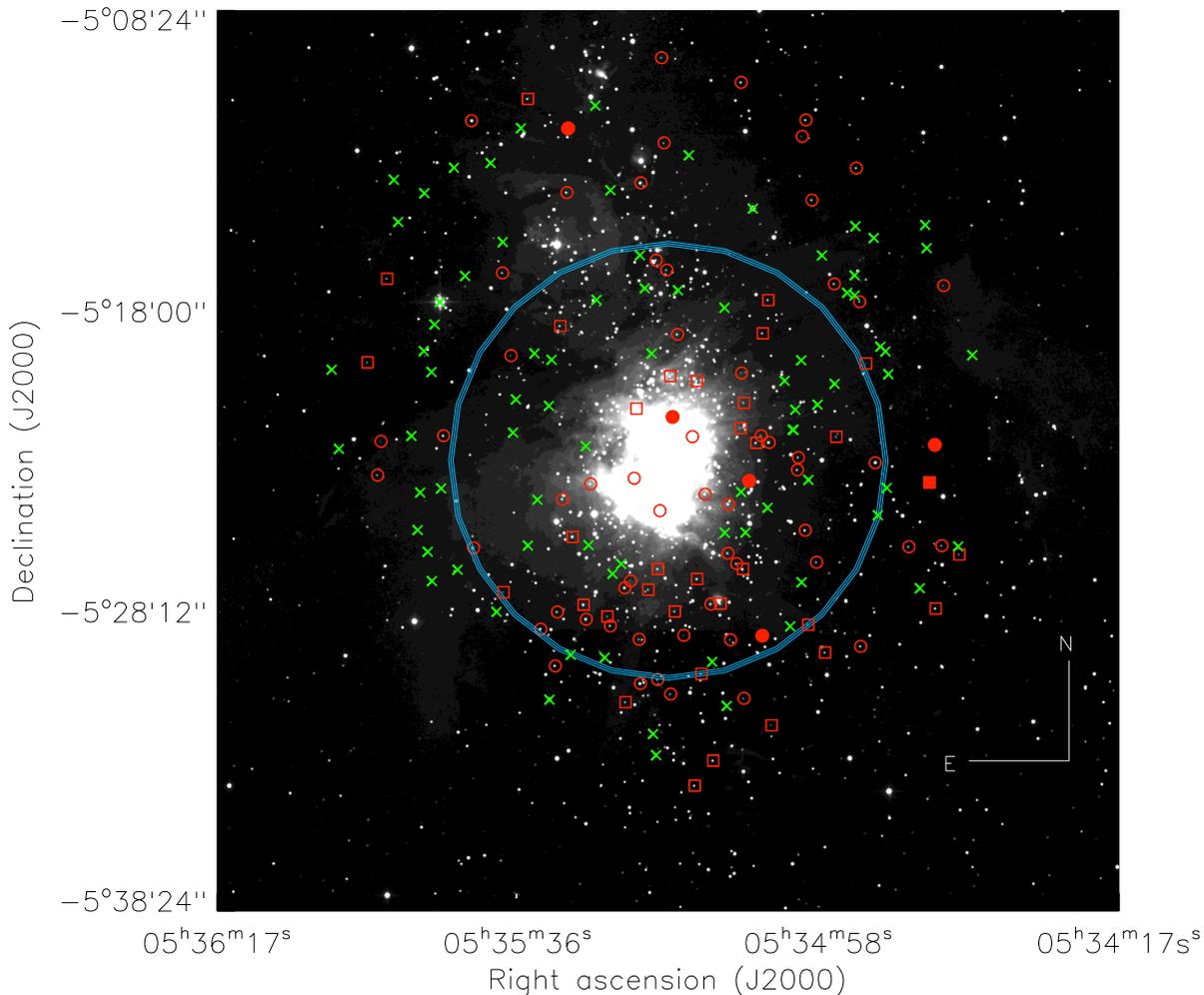}
	\vspace{.9cm}
       \caption{$H$-band {\it 2MASS} image showing the spatial distribution of our sample. The field is centered on Trapezium cluster and 
       covers an area of about 0.5\degr$\times$0.5\degr. Circles and squares show the fibers allocated to stars with $v\sin i$ larger 
       and smaller than 9.0 km s$^{-1}$ (our detection limit; cf. Sect.~\ref{sec:vsini_upper_limit}), respectively. Crosses indicate sky 
       position to where sky fibers were allocated. Filled symbols mark the position of our most probable binary stars. The circle 
       in the center of the field represents the cluster radius. It is centered on $\Theta^1$~Ori and has a radius of about 1 pc 
       (cf. Sect.~\ref{sec:breakup_vel}).}
       \label{fig:sample}
\end{figure*}

\begin{figure}[h]
  \resizebox{\hsize}{!}{\includegraphics{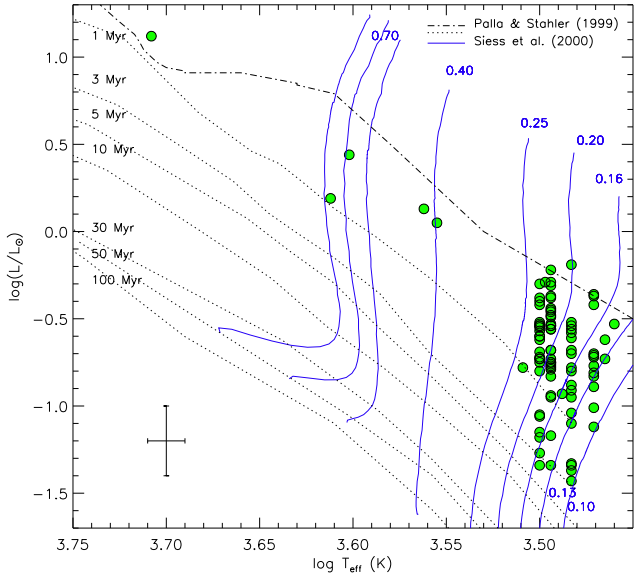}}
       \caption{Positions on the HR diagram of the stars included in our sample. Luminosities and effective temperatures (and 
       their mean errors) are taken from \cite{hillenbrand97}. The \cite{siessetal2000} pre-main sequence evolutionary tracks at 
       X=0.703, Y=0.277, and Z=0.020 are also displayed with the labels representing their masses. The \cite{pallastahler99} 
       birth-line and isochrones (from 1 to 100 Myr) are showed with dash-dotted and dotted lines, respectively. }
       \label{fig:hr}
\end{figure}

This project has been executed as part of the Science Verification of the Fiber Large Array Multi-Element Spectrograph 
({\sc flames}; \citealt{pasquinietal02}) attached to the Kueyen Telescope (UT2) at the Paranal Observatory (ESO). Additional 
observations were obtained with {\sc flames} GTO observations in Period 76A. Both observing runs were performed 
with the multi-object GIRAFFE spectrograph in MEDUSA mode\footnote{This is the observing mode in {\sc flames} in which 
132 fibres with a projected diameter on the sky of 1\farcs{2} feed the GIRAFFE spectrograph. Some fibres are set on the 
target stars and others on the sky background.}. We selected from \cite{hillenbrand97} 96 stars with a membership probability 
given by \cite{joneswalker88} higher than 95\%. Their spatial location is shown in Fig.~\ref{fig:sample}, while their position 
in the HR diagram is shown in Fig.~\ref{fig:hr}. Five stars in our sample, namely, JW50, JW99, JW239, JW669, and JW961 following 
the \cite{joneswalker88} number, have a mass higher than $~0.25M_\odot$ and $\log T_{\rm eff}>3.5$ K. In the end, 91 
targets were selected covering a mass range of $~0.10-0.25M_\odot$ according to \cite{siessetal2000}, i.e. 35\% of the 
\cite{hillenbrand97} sample with $M\le0.25M_\odot$ (cf. Sect.~\ref{sec:disklocking}). 

Observations were carried out in 2003 from January 25th to February 3rd (5 hours in 5 nights) and in 2005--2006 from October 
15th to January 20th (8.4 hours in 11 nights). Two different GIRAFFE setups were chosen. The first setup, HR14 (resolution $R=28\,300$) 
covers the range 638.3--662.6 nm containing the H$\alpha$ line. The second one, HR15 ($R=19\,300$), covers the range 659.9--695.5 nm 
containing the \ion{Li}{i} line at 6708 \AA. A log of the observations is given in Table~\ref{tab:observations}. In particular, 5 
stars (namely JW50, JW99, JW239, JW669, and JW961) were observed only with the lithium set-up.

The GIRAFFE Base-Line Data Reduction Software (girBLDRS\footnote{http://girbldrs.sourceforge.net/.}; \citealt{blecha2000}) 
was used to reduce the data. Sky subtraction is described in Sect.~\ref{sec:Halpha_line}.

\begin{table}[h]  
\caption{Log of the observations.}
\label{tab:observations}
\begin{center}  
\begin{tabular}{lccccc}
\hline
\hline
$\alpha$    & $\delta$  &  Date      &  UT      & $t_{\rm exp}$ & Filter \\
(\degr)     &  (\degr)  & (d/m/y)    & (h:m:s)  &  (s)          &        \\ 
\hline
83.82747      & $-$31.2880   & 25/01/2003 & 02:45:19 & 3600  & HR14   \\
83.82750      & $-$31.2880   & 26/01/2003 & 02:14:47 & 3600  & HR15   \\
83.82762      & $-$31.2881   & 28/01/2003 & 02:32:05 & 3600  & HR15	 \\
83.82766      & $-$31.2881   & 29/01/2003 & 02:10:41 & 3600  & HR15	 \\
83.82763      & $-$31.2880   & 03/02/2003 & 02:35:43 & 3600  & HR15   \\
83.82731      & $-$31.2880   & 15/10/2005 & 07:34:36 & 2820  & HR15   \\
83.82730      & $-$31.2881   & 16/10/2005 & 07:21:02 & 2820  & HR15	 \\
83.82730      & $-$31.2881   & 17/10/2005 & 07:23:30 & 2820  & HR15	 \\
83.82739      & $-$31.2880   & 18/10/2005 & 06:48:58 & 2820  & HR15   \\
83.82729      & $-$31.2880   & 19/10/2005 & 07:23:02 & 2820  & HR15   \\
83.82725      & $-$31.2881   & 20/10/2005 & 07:17:21 & 2820  & HR15	 \\
83.82730      & $-$31.2881   & 21/10/2005 & 06:43:18 & 2820  & HR15	 \\
83.82731      & $-$31.2880   & 04/11/2005 & 06:25:09 & 2820  & HR15   \\
83.82747      & $-$31.2880   & 05/11/2005 & 05:05:07 & 2820  & HR15   \\
83.82718      & $-$31.2881   & 18/01/2006 & 03:28:53 & 2820  & HR15	 \\
83.82707      & $-$31.2881   & 20/01/2006 & 04:05:22 & 2054  & HR15	 \\
\hline
\end{tabular}
\end{center}
\end{table}  

\subsection{Rotational and Radial Velocity Analysis}
Radial velocities ($V_{\rm rad}$) and projected rotational velocities ($v\sin i$) were derived using a cross-correlation
of the object spectrum from 660.6 to 696.5 nm against a CORAVEL-type numerical mask (\citealt{baranneetal79, benzmayor84}) 
based on a M4 stellar spectrum (\citealt{delfosseetal98}). Regions containing telluric lines and strong photospheric lines 
(such as the hydrogen lines) were removed from the mask. The cross-correlation function (CCF) can be fairly well fitted 
by a Gaussian function where the abscissa of its minimum gives the radial velocity and its width, $\sigma_{\rm CCF}$, is 
related to the broadening mechanisms intrinsic to the stars affecting the photospheric lines convolved by the instrumental profile. 

In order to compute $V\sin i$ using $\sigma_{\rm CCF}$, the constants $A$, and $\sigma_0$ need to be determined in 
the expression below:

\begin{equation}
\label{eq:calibration}
v\sin i=A \sqrt{\sigma_{\rm CCF}^2-\sigma_0^2}\,.
\end{equation}

\begin{figure}
\resizebox{\hsize}{!}{\includegraphics{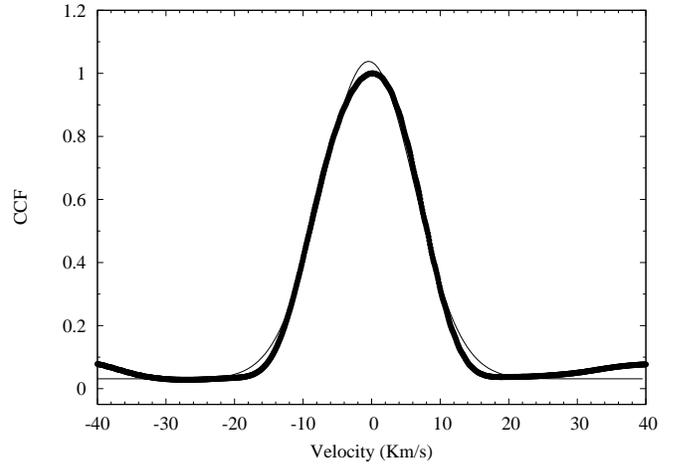}}
\caption{CCF of the Th-Ar emission-line spectrum against a Th-Ar template. The fit gives a $FWHM$ of 15.6 km s$^{-1}$.}
\label{fig:ccf_thar}
\end{figure}

\noindent{The constant $A$ coupling the differential broadening of the CCF to the $v\sin i$ was determined as described in 
\cite{meloetal01}, and has a similar value, 1.9, as found by these authors. The width of the CCF of 
a non-rotating star, $\sigma_0$, however could not be determined as in \cite{meloetal01} due to the lack of calibrators with 
well determined rotational velocities. Instead the $\sigma_{\rm CCF}$ of the Th-Ar spectrum was used as a proxy for 
$\sigma_0$ (Fig.~\ref{fig:ccf_thar}). The caveat is that this determination of $\sigma_0$ includes solely the instrumental 
profile, which is the dominant contributor in the case of low- and mid-resolution spectrographs such as GIRAFFE.}

As an example, the CCF for the {\sc flames} solar spectrum acquired with the same set-up and fiber mode as the stars studied 
here was computed \footnote{Solar spectra for all the GIRAFFE fiber modes and set-ups are available at the web site 
http://www.eso.org/observing/dfo/quality/GIRAFFE\\ /pipeline/solar.html.}. Assuming a $v\sin i=2$ km s$^{-1}$ for the Sun, 
we found a $\sigma_0=7.1$ km s$^{-1}$. Such a difference in the $\sigma_0$ would translate to a difference of 
1.2 km s$^{-1}$ in $(v\sin i)_{\rm min}$ (i.e. the minimum $v\sin i$ we can measure with this GIRAFFE set-up; see cf. 
Sect.~\ref{sec:vsini_upper_limit}). Therefore, we adopt $A=1.9$ and $\sigma_0=6.62$ km s$^{-1}$ as the constants in 
Eq.~\ref{eq:calibration}.

\subsection{$v\sin i$ and $V_{\rm rad}$ errors}
Photon noise errors were estimated as in \cite{meloetal01}. The procedure is summarized as follows. A grid of artificially 
rotated spectra from 5 to 20 km s$^{-1}$ was created based on the GIRAFFE solar spectrum. Gaussian noise was added to 
simulate a $S/N$ varying from 5 to 40. For each point on the grid, 96 spectra were generated. The error of $V_{\rm rad}$ is 
similar to the error of $v\sin i$, which is given by:
 
\begin{equation}
\sigma_{v\sin i} \sim \sigma_{V_{\rm rad}} = \frac{1.4 \times v\sin i}{S/N} - 0.25\,.
\label{eq:sigma_v}
\end{equation}
 
For almost all targets several measurements are available. In such cases, the errors quoted in Table~\ref{tab:sample} are 
simply based on the scatter of the individual measurements. For four stars (namely, JW99, JW689, JW964, and JW1032) we 
obtained only one good measurement. In such cases, the errors quoted in Table~\ref{tab:sample} were computed from the Eq.~\ref{eq:sigma_v}.

\subsection{$v\sin i$ detection limit}
\label{sec:vsini_upper_limit}
The smallest $v\sin i$ that can be measured was estimated using Eq. 4 of \cite{meloetal01}. Visual inspection of 
different lines of Th-Ar shows that the typical $\Delta \sigma_0$ is of the order of $1$ km s$^{-1}$. To 
this value, one has to add the typical (external) error ($\sigma_{\rm CCF}^{\rm ext}$) of the measurement of 
$\sigma_{\rm CCF}$ itself. For stars in our sample with multiple observations the mean rms of $\sigma_{CCF}$ is about 0.7 km s$^{-1}$.

Plugging these values into Eq. 4 of \cite{meloetal01}, we find: 
$(v\sin i)_{\rm min} \la A\sqrt{2 \sigma_0 \epsilon} \sim 9.0$ km s$^{-1}$, 
where $A=1.9$, $\sigma_0=6.62$ km s$^{-1}$, and $\epsilon=\Delta \sigma_0+\sigma_{\rm CCF}^{\rm ext}=1.7$ km s$^{-1}$.

\subsection{Disk and accretion diagnostics}
\subsubsection{$\Delta(I_{\rm C}-K)$ excess}
The $I_{\rm C}$-band fluxes are least affected by circumstellar activity and typically dominated by photospheric emission, 
while emission arising from accretion disks is maximized at $K$-band. \cite{hillenbrandetal98} defined a quantity that they used 
to measure the magnitude of the near-infrared excess:
\begin{equation}
\Delta(I_{\rm C}-K)=(I_{\rm C}-K)_{\rm observed}-0.5A_V-(I_{\rm C}-K)_{\rm photosphere} 
\end{equation}
\noindent{which represent the difference between the observed $(I_{\rm C}-K)$ color, and the contributions of the reddening and 
the underlying stellar photosphere. The principal uncertainty in this parameter comes from the photometric error and the time variability 
of the photometry (\citealt{hillenbrandetal98}). These authors adopted a value of $\Delta(I_{\rm C}-K)=0.3$ to divide the 
disk-less stars from the disked ones.}

\subsubsection{H$\alpha$ line}
\label{sec:Halpha_line}
The width of H$\alpha$ profiles is commonly used as an indicator of accretion in T-Tauri stars. Accreting T-Tauri stars (CTTS) 
show strong, broad H$\alpha$ profiles whose origin is related to infall of high-velocity material forming their circumstellar 
disk onto the photosphere (e.g., \citealt{hartmann94}), while weak-line T-Tauri stars (WTTS) show a much narrower H$\alpha$ profile, 
mostly produced by strong chromospheric activity. Due to differences in the continuum level in the region close to the H$\alpha$ line 
for different temperatures, no unique value of its equivalent width can be used to distinguish between CTTS and WTTS (\citealt{martin98}). 
Hence, the threshold for classifying an object as an accretor depends on the spectral type. \cite{whitebasri03} 
proposed a full width of H$\alpha$ at 10\% of the line's peak intensity (what they called the 
10\% width) as a more robust and unified criterion to diagnose stellar accretion regardless the spectral type. Using the presence 
of veiling as an accretion criterion, they proposed that a 10\% width larger than 270 km s$^{-1}$ indicates the 
star is accreting, and therefore can be classified as a CTTS. Using physical reasons and empirical findings, 
\cite{jayawardhanaetal2003} adopted 200 km s$^{-1}$ as the accretion cutoff for the very low-mass regime 
(down to the limit of hydrogen burning). This value corresponds to a mass accretion rate of about $10^{-10} M_\odot$ yr$^{-1}$. 
Measurements of 10\% widths are advantageous over optical veiling and H$\alpha$ equivalent width measurements because 
they can be extracted over a short wavelength range, do not depend on the underlying stellar luminosity, and do not depend 
on the availability of a comparison template (\citealt{whitebasri03}).  

\begin{figure*}[t]
 \begin{center}
  \resizebox{\hsize}{!}{\includegraphics{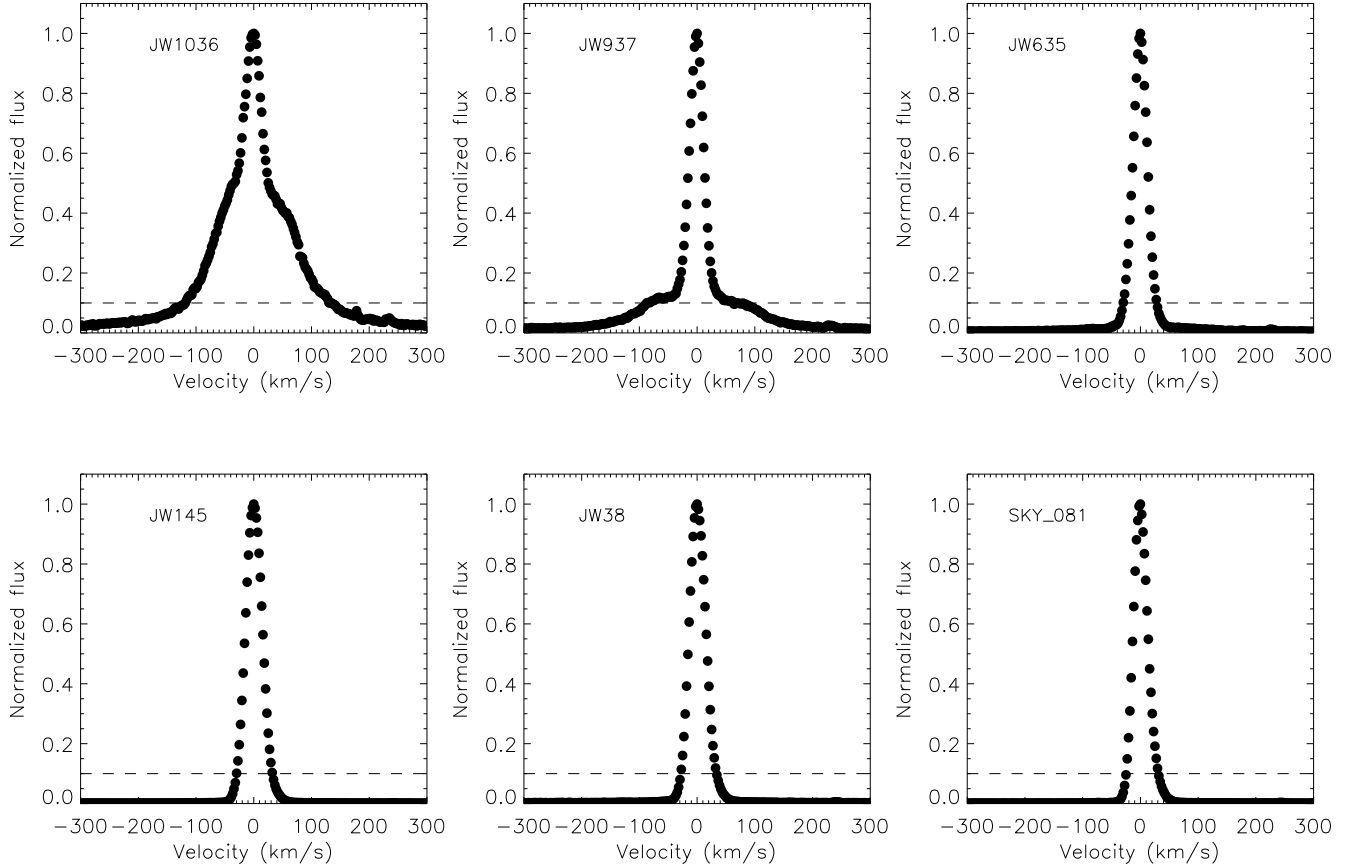}}
    \caption{H$\alpha$ velocity profiles for five different objects of the ONC and one sky position. The H$\alpha$ 10\% 
    intensity is showed with a dashed line.}
    \label{fig:10_percent_spectra}
  \end{center}
\end{figure*}

In order to be able to compute the 10\% widths for all fibers (i.e., stellar and sky spectra), we first fitted and then subtracted the 
continuum around 654 nm up to 660 nm. The continuum subtracted spectra were then normalized by their highest intensity. 
Some examples of the final spectra used to computed the 10\% widths are shown in Fig.~\ref{fig:10_percent_spectra}.

\begin{figure}	%[t]
  \resizebox{\hsize}{!}{\includegraphics{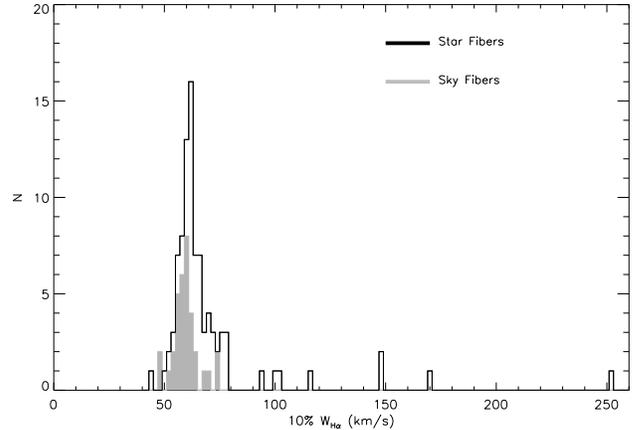}}
         \caption{Distribution of 10\% widths. Sky emission distribution is shown with the filled grey histogram whereas the objects are 
	 displayed in continuous black line. Stars showing 10\% widths below $\sim 70$ km s$^{-1}$ are likely to be dominated by the
       sky emission.}
       \label{fig:eqw10}
\end{figure}

The H$\alpha$ profiles in the ONC are known to be strongly contaminated by the nebular emission. However, at our spectral resolution, 
the sky emission lines are much narrower than the profile expected for an accreting star (\citealt{stassunetal99}). As an 
additional check of the level of contamination, we compared the 10\% widths measured for the object fibers and those allocated to 
sky positions. As we can see in Fig.~\ref{fig:eqw10}, sky fibers show a rather narrow distribution of 10\% widths in the range of 
50-70 km s$^{-1}$. In contrast, stars have values from 50 to 250 km s$^{-1}$. It is worth noticing that, according to 
\cite{whitebasri03} criterion, none of the ONC stars studied here is accreting. This will be better discussed in Sect.~\ref{sec:summ_concl}.

\subsubsection{[3.6]$-$[8.0] $\mu$m flux}
The Orion Molecular Cloud was surveyed with the {\it Spitzer Space Telescope} as part of GTO program (\citealt{megeathetal2008}). 
The IRAC@{\it Spitzer} mid-infrered bands can be used as circumstellar disk diagnostics (\citealt{rebulletal2006}). In particular, 
the color index based on the two most widely separated bands, namely 3.6 $\mu$m and 8 $\mu$m, 
is a very useful method for distinguishing disk and non-disk candidates, with [3.6]$-$[8.0]=1 the boundary between disk 
and non-disk candidates (\citealt{rebulletal2006}).

\subsubsection{\ion{Ca}{ii} infrared triplet}
As pointed out by \cite{hillenbrandetal98}, the \ion{Ca}{ii} infrared triplet ($\lambda\lambda$8498, 8542, 8662) can 
be used to explore the presence of accreting disks. The triplet emission features exhibit narrow and broad components, which seem 
to show a correlation with spectral veiling measured at optical wavelengths, implying a correlation with 
the mass-accretion rate. The authors analyze the behavior of the equivalent widths (EWs) of the strongest level (i.e. $\lambda$8542) 
in their sample of low-mass ONC stars. According to their convention, stars are supposed to have accretion disks if the sum of 
the EWs of the broad and narrow components is filled or in emission ($W_{\rm CaII}=0\pm1$ \AA). They emphasize that 
definite conclusions regarding either disk frequency or accretion 
properties based on the \ion{Ca}{ii} infrared lines await studies capable of separating photospheric absorption, 
narrow-component emission, and broad-component emission. In Table~\ref{tab:sample} the column $W_{\ion{Ca}{ii}}$ 
lists the EWs of our targets taken from \cite{hillenbrandetal98}, where values of 0.0 indicate that 
neither emission nor absorption features were apparent from visual inspection of the spectrum. 

\section{Results and discussion}

\subsection{Radial velocities and spectroscopic binaries}
\label{sec:rad_vel}
\begin{figure}[t]
  \resizebox{\hsize}{!}{\includegraphics{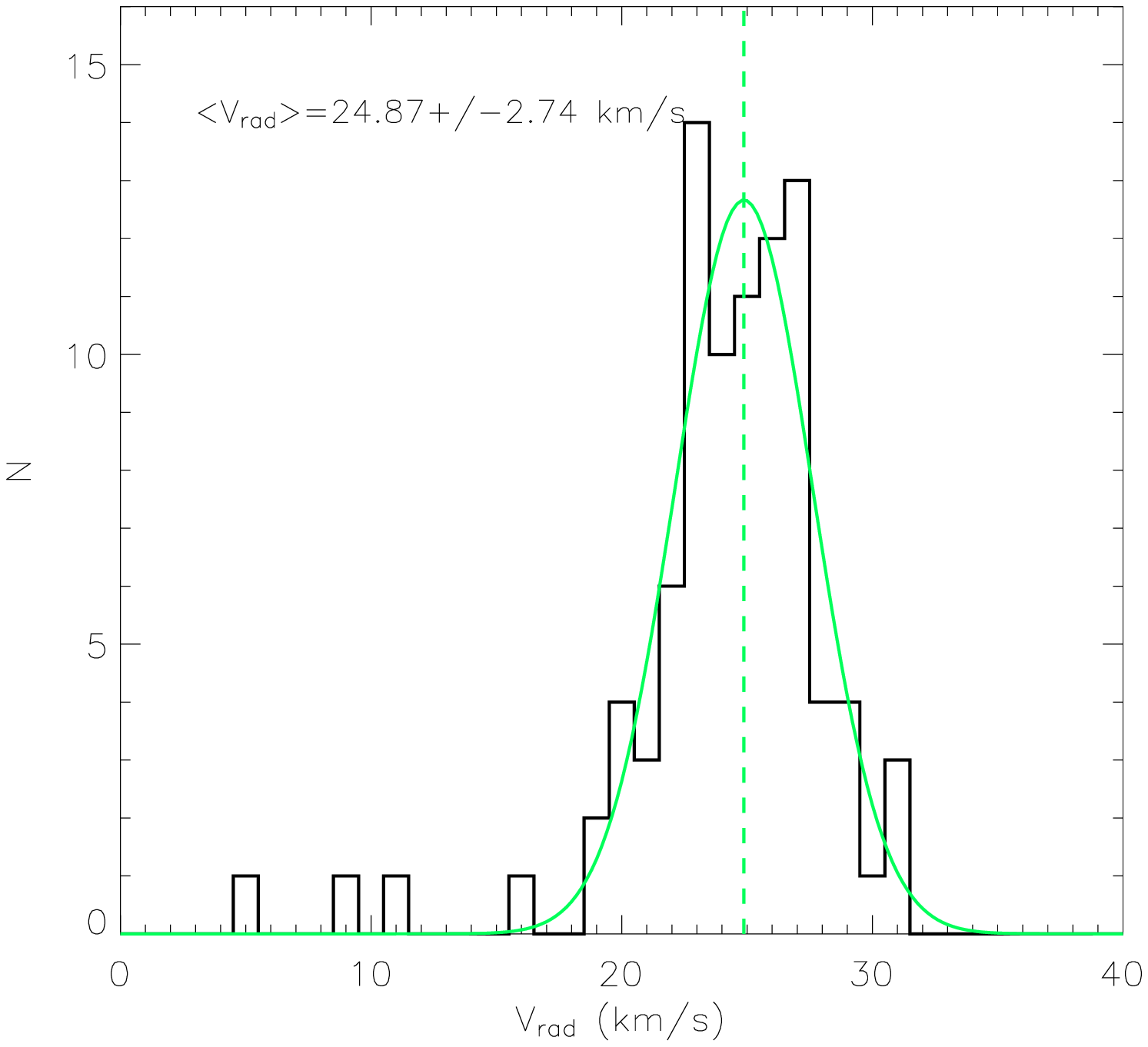}}
       \caption{Radial velocity distribution for our sample of stars with $M<0.25M_{\odot}$. For the most probable binaries 
       we plot the mean values of the two components (Table~\ref{tab:sample}). The Gaussian fit of the distribution yields a 
       mean $V_{\rm rad}$ of 24.87 km s$^{-1}$ (with a $\sigma_V{}_{\rm rad}$ of 2.74 km s$^{-1}$), which is fully consistent 
       with the membership in the ONC (e.g., \citealt{stassunetal99}).}
       \label{fig:vr_histo}
\end{figure}

A histogram of radial velocity measurements for the sample of low-mass stars is shown in Fig.~\ref{fig:vr_histo}. 
The Gaussian fit of the distribution yields a mean $V_{\rm rad}$ of 24.87 km s$^{-1}$ and $\sigma_V{}_{\rm rad}$ of 2.74 km s$^{-1}$ 
in very good agreement with the typical values found in the literature (see, e.g., \citealt{stassunetal99,alcalaetal00,covinoetal01,rhodeetal01}). 
Our radial velocities confirm the membership of all stars in the sample, as expected, since the targets have been 
selected based on the membership probability given by \cite{joneswalker88}.

\begin{figure*}[t]
  \begin{center}
  \includegraphics[width=5.9cm,height=5.9cm]{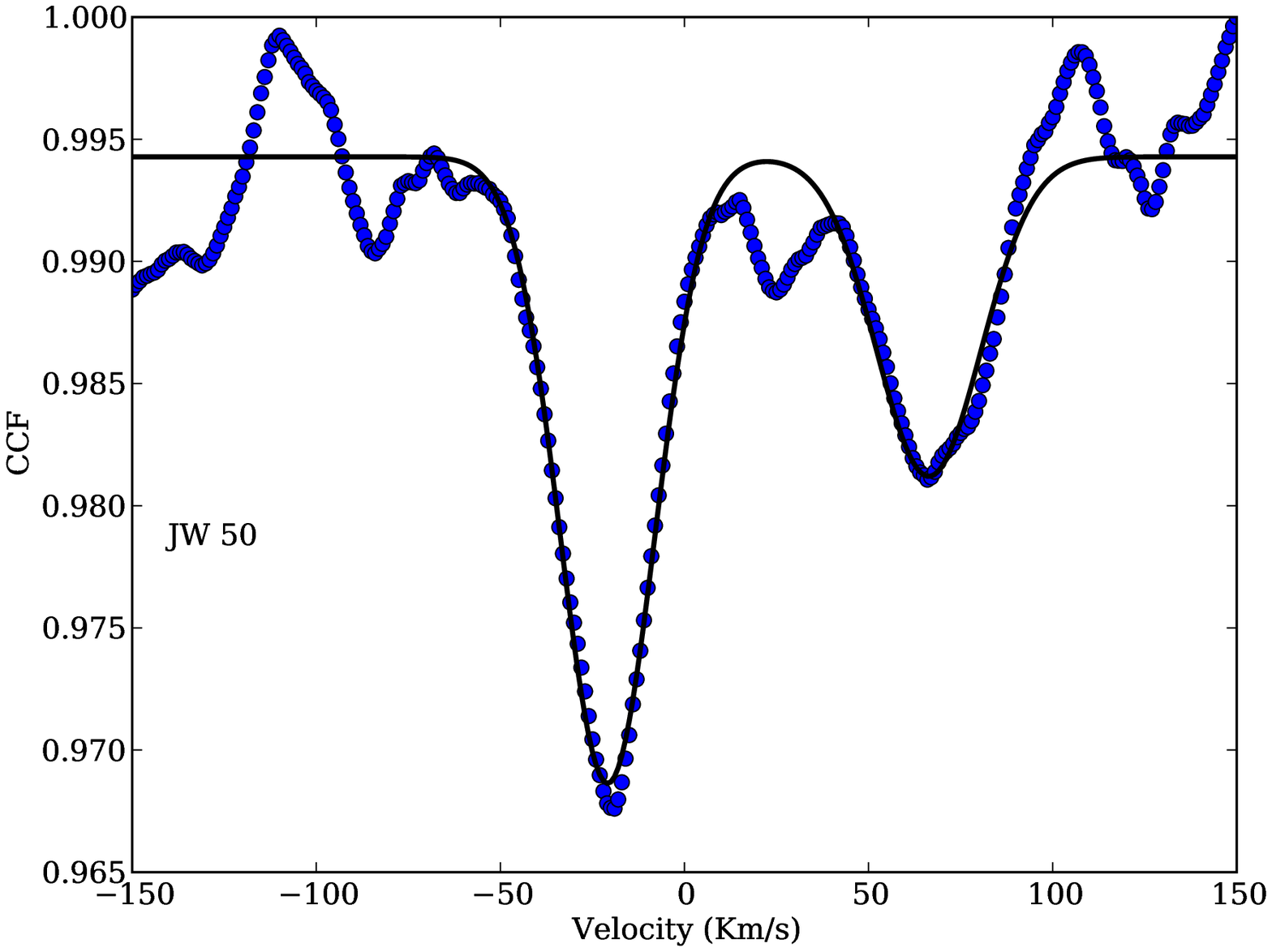}
  \includegraphics[width=5.9cm,height=5.9cm]{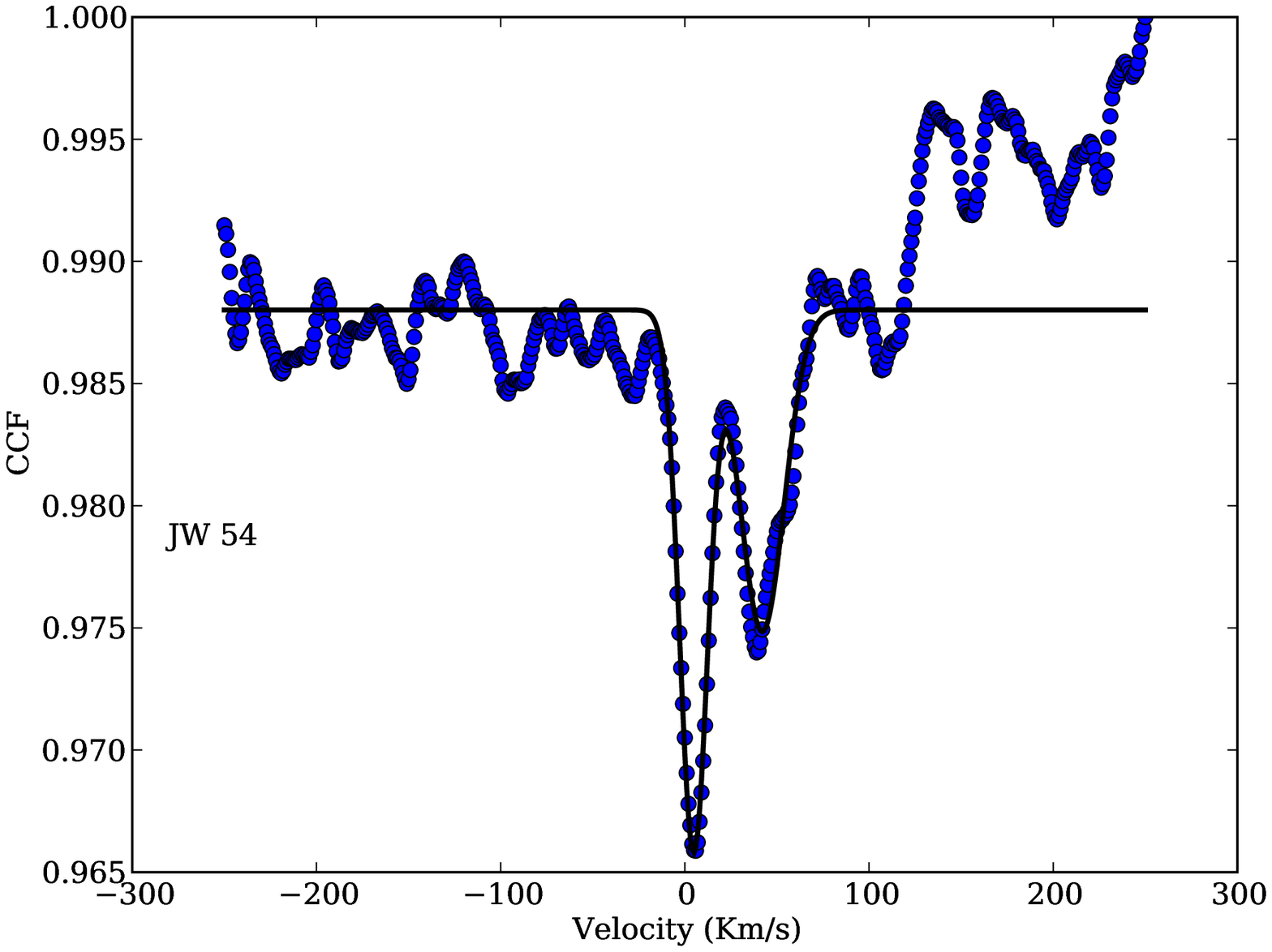}
  \includegraphics[width=5.9cm,height=5.9cm]{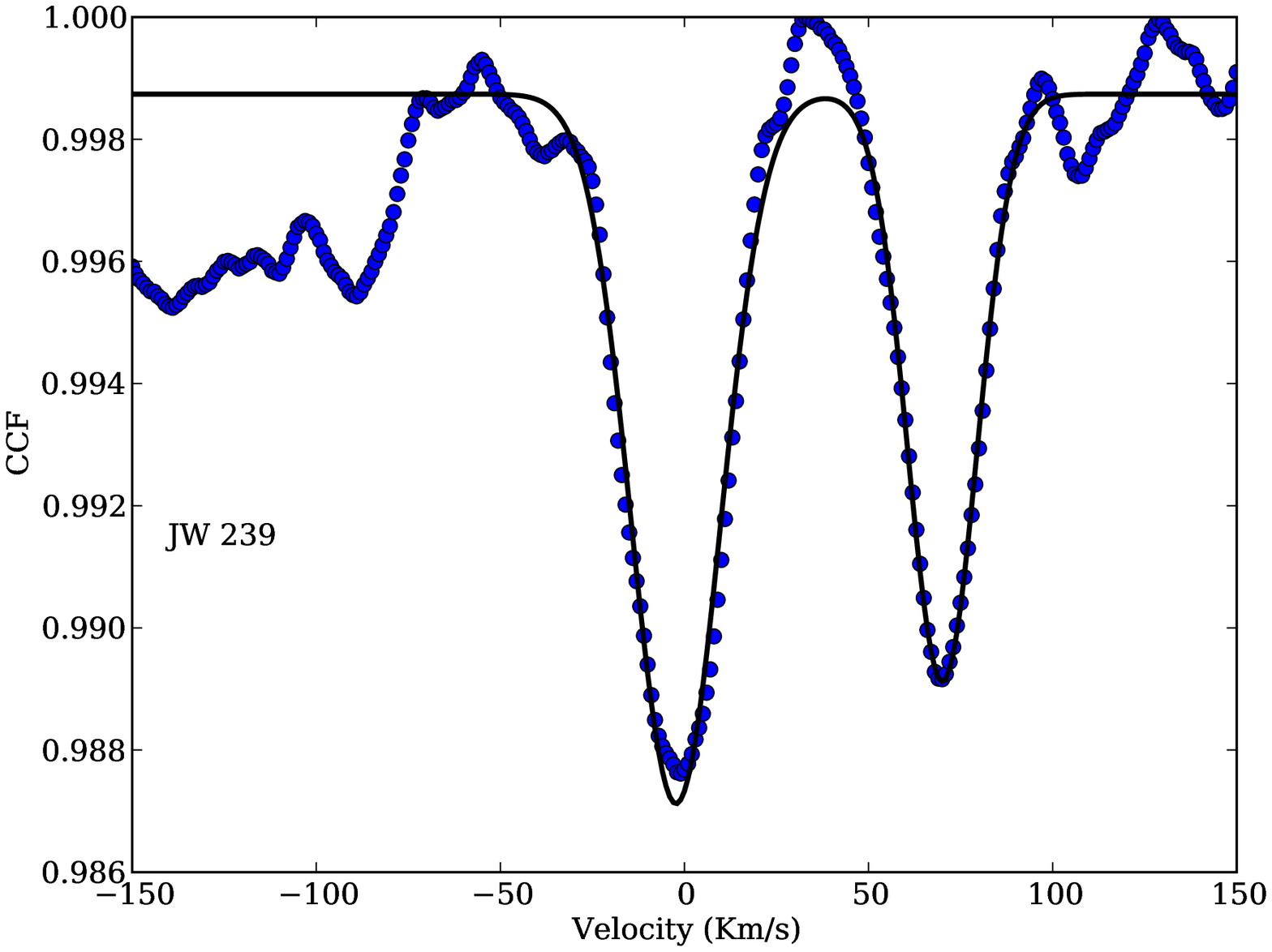}
  \includegraphics[width=5.9cm,height=5.9cm]{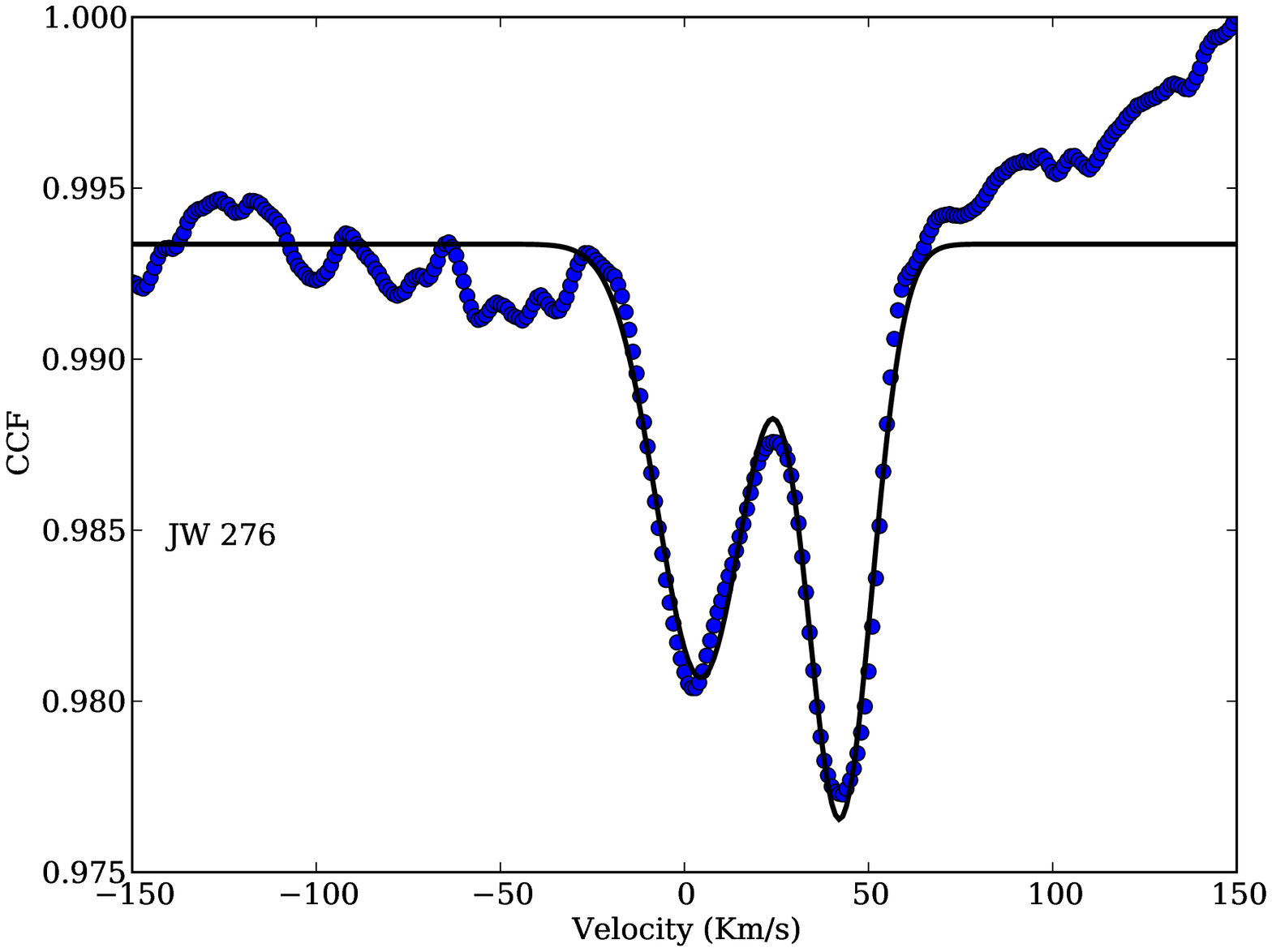}
  \includegraphics[width=5.9cm,height=5.9cm]{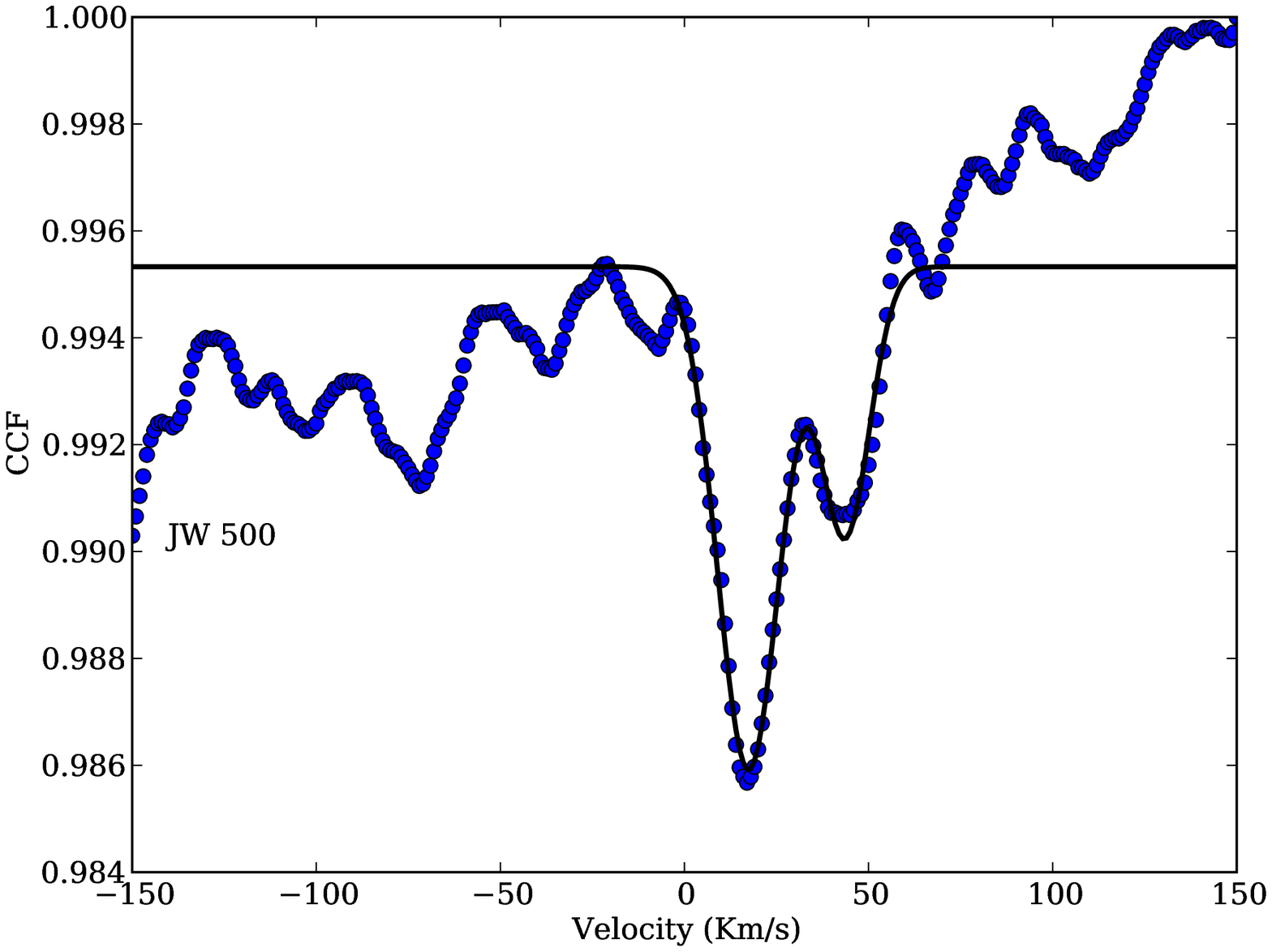}
  \includegraphics[width=5.9cm,height=5.9cm]{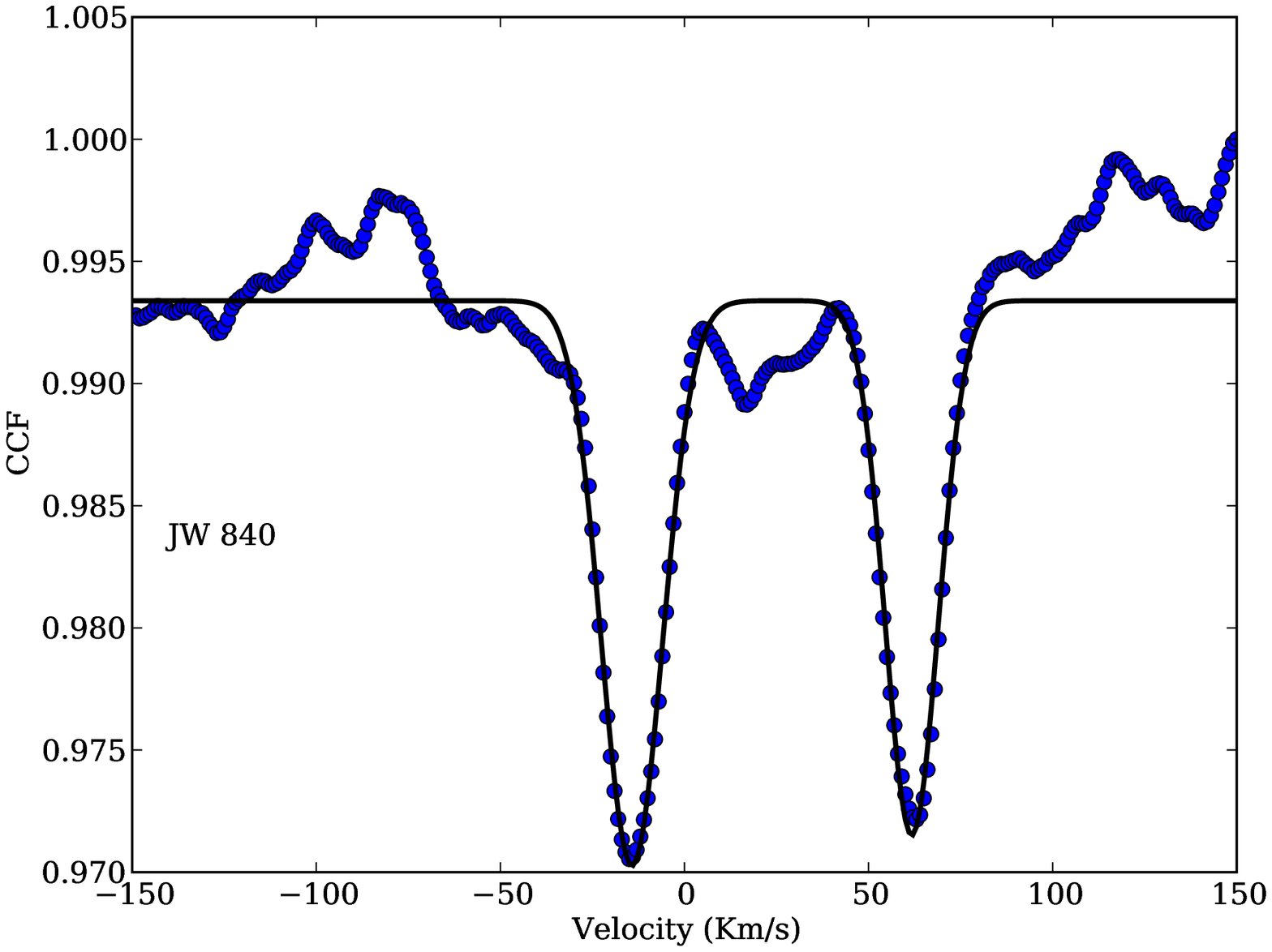}
  \end{center}
       \caption{Double-lined CCF profiles for our most probable binary stars.}
       \label{fig:CCF_binaries}
\end{figure*}

Among the 96 stars observed, 6 (JW50, JW54, JW239, JW276, JW500, and JW840) show a double-lined CCF (Fig.~\ref{fig:CCF_binaries}). 
Three of these, JW50, JW239, and JW840, were already classified as binaries by \cite{joneswalker88} and \cite{tobinetal2009}, 
and two of them (namely JW50 and JW239) have $M>0.25M_{\odot}$ (Fig.~\ref{fig:hr}). For our most probable low-mass binaries we show in 
Fig.~\ref{fig:vr_histo} an average of the radial velocities of the two components. The $V{}_{\rm rad}$ values for both the components 
and their mean values are also listed in Table~\ref{tab:sample} for all the sample. 

Future radial velocity follow-up of such targets would help to solve their 
orbits and to determine the circularization period in this mass range which can bring interesting information about the frequency 
of spectroscopic binaries and internal dissipation mechanisms of angular momentum related to the convective motion within the stars 
(e.g., \citealt{meloetal01b}).

\subsection{Rotation period and radius from $v\sin i$}
\begin{figure}[t]
  \resizebox{\hsize}{!}{\includegraphics{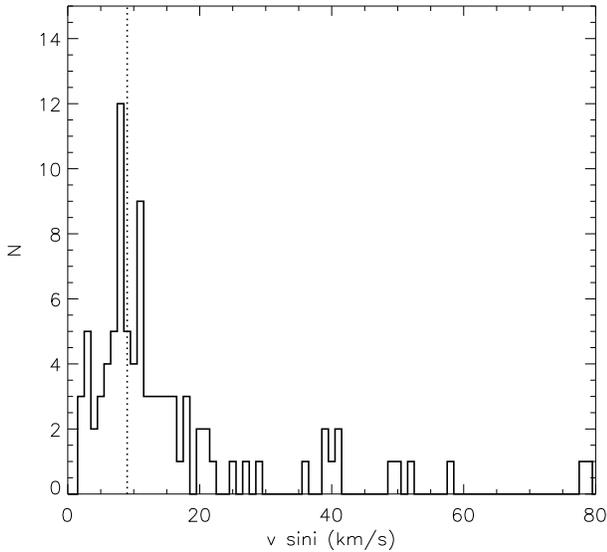}}
       \caption{$v\sin i$ distribution for our low-mass stars. Our $v\sin i$ limit of 9 km s$^{-1}$ is shown with a dotted line.}
       \label{fig:vsini_distr}
\end{figure}

\begin{figure}[t]
  \resizebox{\hsize}{!}{\includegraphics{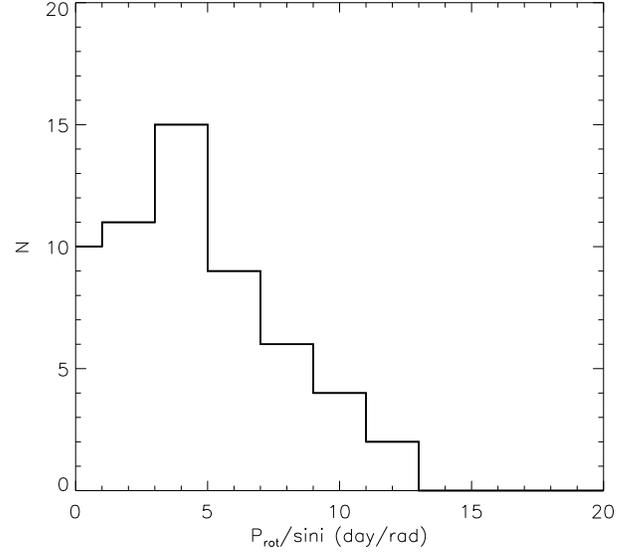}}
       \caption{Rotational period distribution obtained with our $v\sin i>9$ km s$^{-1}$ values (namely, 57).}
       \label{fig:prot_distr}
\end{figure}

We considered the low-mass stars having both $v\sin i$ (Fig.~\ref{fig:vsini_distr}) above our detection limit of 9 km s$^{-1}$ as well 
as radii determined by \cite{hillenbrand97}. For these 57 objects we computed $P_{\rm rot}/\sin i=2\pi R/v \sin i$. For the binary stars, 
we considered $v\sin i$ values of the first component listed in Table~\ref{tab:sample}. These periods distribute according to 
a unimodal distribution peaked around $P_{\rm rot}/\sin i=3-5$ day/rad (Fig.~\ref{fig:prot_distr}). Considering a statistical 
correction of $<\sin i>=\pi/4$ (cf. Sect.~\ref{sec:vsini_veq}), our period distribution seems to have a peak at $\sim2.4-3.9$ days, 
while with our $<\sin i>=0.61$ (cf. Sect.~\ref{sec:vsini_veq}) the peak is around $1.8-3.1$ days. This result is in agreement with previous 
ones present in the literature for stars with $M<0.25M_\odot$ (see, e.g., \citealt{Herbstetal2007}, and references therein). Namely, 
whereas stars with $M>0.25 M_\odot$ present a bimodal rotational period distribution with two peaks near 2 and 8 days, low-mass stars 
($M<0.25M_\odot$) show a unimodal distribution peaked near 2 days. 

Combining our $v\sin i>9$ km s$^{-1}$ values with the rotation period measured by \cite{herbstetal02} we can constrain the minimum stellar 
radii for 23 low-mass stars. Taking into account the relationship $R \sin i=P_{\rm rot} v\sin i/2\pi$, we find our $R \sin i$ ranges from 
0.7 to 1.2 $R_{\odot}\cdot$rad, with a distribution peaked to around $0.9-1.5 R_{\odot}$ if we consider $<\sin i>=\pi/4$, and 
$\sim 1.1-2.0 R_{\odot}$ for our $<\sin i>=0.61$ (cf. Sect.~\ref{sec:vsini_veq}). The \cite{hillenbrand97}'s radii for the same stars 
show a peak around $1.3-1.4 R_{\odot}$.

\subsection{$v\sin i$ and $v_{\rm eq}$}
\label{sec:vsini_veq}
\begin{figure}[t]
  \resizebox{\hsize}{!}{\includegraphics{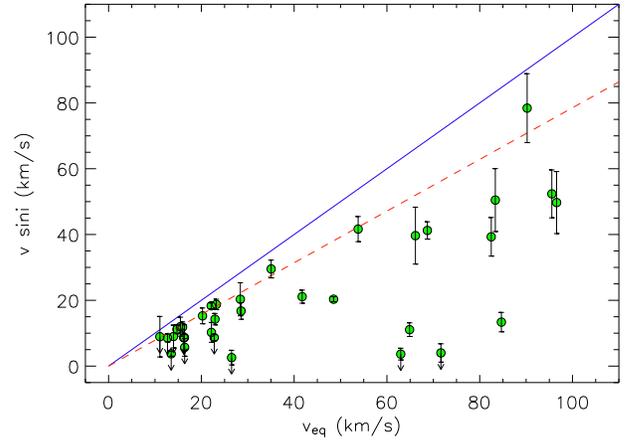}}
       \caption{$v\sin i$ versus $v_{\rm eq}$. The continuous and dashed lines represent the equations $v\sin i=v_{\rm eq}$ 
       and $v\sin i=(\pi/4)v_{\rm eq}$, respectively. The arrows refer to our detection limit of 9 km s$^{-1}$.}
       \label{fig:vsini_veq}
\end{figure}

We compared $v\sin i$ with equatorial velocities $v_{\rm eq}=2\pi R/P_{\rm rot}$ calculated adopting the radii and rotational 
periods by \cite{hillenbrand97} and \cite{herbstetal02}, respectively (Fig.~\ref{fig:vsini_veq}). For stars with $v\sin i<9$ 
km s$^{-1}$ detection limit arrows are plotted. The solid and dashed lines mark $v\sin i=v_{\rm eq}$ and $v\sin i=(\pi/4)v_{\rm eq}$, 
respectively. Similar to what was found by \cite{rhodeetal01}, almost all our 34 stars fall below the $v\sin i=(\pi/4)v_{\rm eq}$ 
line, indicating that the average $\sin i$ value we measure is lower than the expected value. We computed the mean $\sin i$ 
considering the stars with $v\sin i>9$ km s$^{-1}$, finding $<\sin i>=0.61\pm0.04$, very close to previous values 
(see, e.g., \citealt{rhodeetal01}, and references therein), but significantly lower than the value of $\pi/4\sim0.785$ expected for a random 
oriented distribution of stellar rotation axes (\citealt{ChadraMunch1950}). This departure has been also observed in previous studies 
of PMS, and a number of explanations, including real physical phenomena, and systematic errors in astrophysical quantities ($v\sin i$, 
$L$, $T_{\rm eff}$, $P_{\rm rot}$) have been invoked by several authors. \cite{rhodeetal01} explored, in their sample of 
$0.1<M<2.5 M_\odot$ stars, all these possibilities finding that the correct $\sin i$ value is produced assuming that the effective 
temperatures have been underestimated by 400--600 K. Trying to resolve this problem is beyond the scope of this work. 
We would only stress that the radii computations of \cite{hillenbrand97} were based on a distance of 470 pc, while 
recent determinations estimate a distance to ONC of 414 pc (\citealt{mentenetal2007}). This leads to $R$, $P_{\rm rot}/\sin i$, 
and $v_{\rm eq}$ 12\% smaller than we find, and a mean $\sin i$ of 0.683, i.e. closer to the statistics value. 
This means our findings are also partially linked to the distance determinations.

Moreover, Fig.~\ref{fig:vsini_veq} shows there is a correlation between $v\sin i$ and $v_{\rm eq}$, demostrating, as pointed 
out by \cite{rhodeetal01}, that the periodicity of our T Tauri stars is caused by the rotation of stars with spotted photospheres.

\subsection{Correlations with $v\sin i$}
Our $v\sin i$ results are given in Table~\ref{tab:sample}. Of the 91 stars observed with $0.10<M<0.25M_{\odot}$, 4 stars are 
{\it bona-fide} spectroscopic binaries. The binary stars will be marked with different symbols in the following plots since tidal effect 
may be influencing their surface rotations. Since the $v\sin i$ values of the binary system components are very similar to each 
other, we decided to plot only the first component listed in Table~\ref{tab:sample}. Finally, 34 stars have $v\sin i$ less 
than or equal to our limit of 9 km s$^{-1}$.

\subsubsection{Are stars rotating close to break-up velocity?}
\label{sec:breakup_vel}
\begin{figure}	%[t]
  \resizebox{\hsize}{!}{\includegraphics{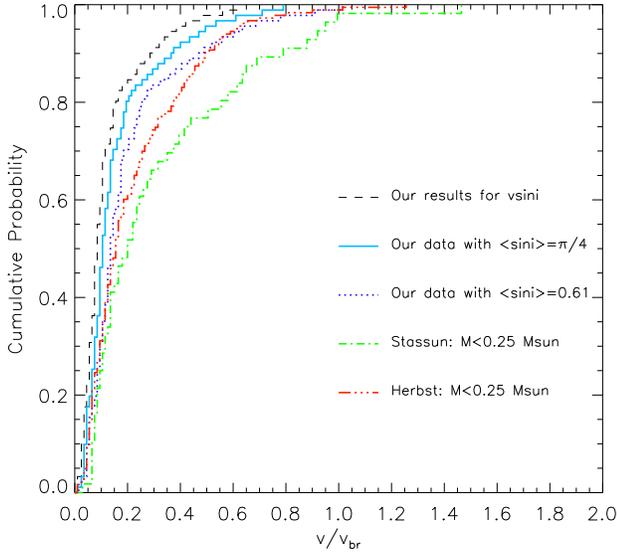}}\\
       \caption{Cumulative distribution of equatorial velocities normalized to break-up velocity. Our sample (dashed, continuous, 
       and dotted lines) is compared to the \cite{stassunetal99} low-mass sample (dash-dotted line) and to the \cite{herbstetal02}'s 
       low-mass sample (dot-dot-dot-dashed line).}
       \label{fig:vbr_histo}
\end{figure}

In Fig.~\ref{fig:vbr_histo} we show the cumulative distribution normalized to the break-up velocity $v_{\rm br}=\sqrt{GM/R}$ 
(where $G$ is the gravitational constant) for our low-mass stars. The break-up velocities from our stars were computed using 
the radii $R$ and masses $M$ taken from \cite{hillenbrand97}. We compare our results with the $v/v_{\rm br}$ values derived 
by \cite{stassunetal99} for $M\le 0.25M_\odot$ and with the values obtained considering the $P_{\rm rot}$ derived by \cite{herbstetal02} 
and mass and radii taken from \cite{hillenbrand97} in the same low-mass regime. 

We show with different lines our results obtained considering $v \sin i$ (without the $\sin i$ correction) and the values 
we obtain for our $\sin i$ mean value of 0.61 and the statistical value of about 0.785 expected for a random oriented 
distribution of stellar rotation axes. A visual inspection suggests that our distributions are very different from the 
\cite{stassunetal99} findings. While 23\% of the 56 \cite{stassunetal99}'s low-mass stars are rotating faster than 
$0.5 v_{\rm br}$, only four stars (namely JW122, JW168, JW530, and JW964) in our sample of 91 stars seems to exceed 
the $0.5 v_{\rm br}$ limit when we consider $<\sin i>=\pi/4$. This number obviously increases to 8 if we consider our $<\sin i>=0.61$. 
Similar results are found for the 183 low-mass stars of the \cite{herbstetal02}'s sample, where the percentage of stars 
with $v/v_{\rm br} \ge 0.5$ is about 10\%. Thus, we can definitely assert that in our sample, no 
star is rotating close to the break-up velocity estimated for its mass and radius. Excluding effects due to mass selection 
effects of the samples, an effect linked to the spatial segregation cannot be excluded, because the stellar population observed 
by \cite{stassunetal99} spans a big area of $40'\times80'$ centered on the Trapezium in ONC, while our stars are localized in an 
smaller area of $20'\times24'$. Also the stars of \cite{herbstetal02} are localized in a smaller area ($33'\times34'$).
In particular, as shown in Fig.~\ref{fig:sample}, about 62\% of our sample is almost inside the circle with radius equal to 
the so-called cluster radius ($R_{\rm cluster}$; \citealt{hillenbrand97}) where \cite{Rodri-Lede2009} find an indication of higher 
$P_{\rm rot}$ compared to the stars outside $R_{\rm cluster}$. This could indicate objects closer to the Trapezium center tend 
to rotate on average slower than the outer ones. An inspection of the \cite{stassunetal99}'s $P_{\rm rot}$ data seems to indicate 
an almost flat distribution for radii smaller than $R_{\rm cluster}$ and the hint of a peak at low $P_{\rm rot}$ for radii bigger 
than $R_{\rm cluster}$. With our smaller sample, we are not able to reach any successful conclusion. We would need more stars and/or 
$v\sin i$ measurements for all their stars. As a consequence, we caution the reader about this issue.

We also computed the Kolmogorov-Smirnov (KS) probability (\citealt{press92}) that the our arrays of data obtained with 
$<\sin i>=\pi/4$ and the \cite{stassunetal99}'s values for low-mass stars are drawn from the same distribution. We find that the 
KS probability that these data were drawn from the same distribution is only of 0.04\%. A value of 0.5\% is obtained 
considering the stars of \cite{stassunetal99} with $M<0.25M_\odot$ locatized in the same spatial region of our sample. This 
comparison shows that, at least as far as our data is concerned, there is not a fast rotator population in the mass 
regime covered by our data. 
 
\subsubsection{Is there any evidence supporting disk-locking?}
\label{sec:disklocking}
One of the well known implications of the disk-locking scenario is that slowly rotating stars should have accreting disks around 
them while rapidly rotating stars should not. In fact, this picture has been supported over the years
in many different papers (e.g., \citealt{edwardsetal93, bouvieretal93, choiherbst96, herbstetal01, rebulletal02}).

In particular, if the diagnostics we use here ($\Delta(I_{\rm C} -K)$ excess, 10\% H$\alpha$ width, [3.8]$-$[8.0], 
and \ion{Ca}{ii} infrared triplet) and rotation are correlated, in the sense that slow rotators are more likely to show excess and 
variability, this is evidence in support of the disk-locking paradigm. As pointed out recently by \cite{IrwinBouvier2009}, 
this means that the population of slowly-rotating stars should show the presence of a disk (e.g., mid-IR excess) or of active 
accretion (i.e. be CTTS), whereas the rapidly-rotating stars should not have disk, or have recently-dissipated disks, and not be 
active accretors (i.e. WTTS). \cite{rebulletal2006}, analyzing {\it Spitzer} mid-IR data for about 900 stars in Orion in the 
mass range $0.1-3 M_\odot$ find that slowly-rotating stars are indeed more likely to posses disks than rapidly-rotating stars. 
However, they also find a puzzling population of slow-rotating stars without disks. This result has been also confirmed subsequently 
(e.g., \citealt{CiezaBaliber2007}).

\begin{figure}	%[b!]
\begin{center}
 \begin{tabular}{c}
  \resizebox{\hsize}{!}{\includegraphics{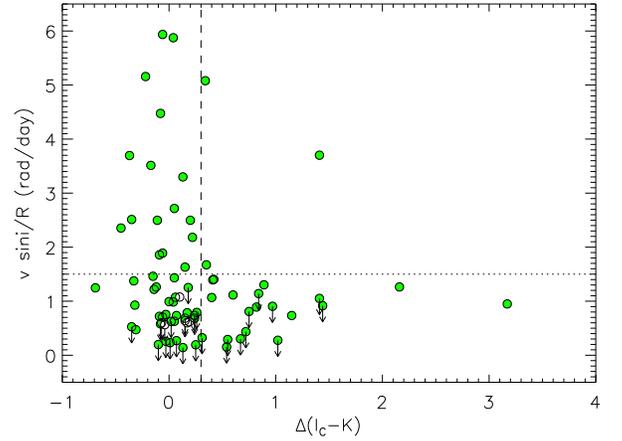}}\\                  
 \end{tabular}
       \caption{Rotation against disk tracers. Rotational velocity as a function of the near infrared excess. A
       drop in the rotation rate is clearly seen for stars with a circumstellar disk (i.e., $\Delta(I_{\rm C}-K)>0.3$). 
       The empty circles represent the position of the most probable binaries. The dotted line marks the 
       division we considered for the Fisher's test.}
       \label{fig:vsini_disk1}
 \end{center}
\end{figure}

In Fig.~\ref{fig:vsini_disk1} we plot our $v\sin i$ measurements divided by the radii as a function
of the near-infrared color excess $\Delta(I_{\rm C}-K)$ given by \cite{hillenbrandetal98} and which is taken as an inner 
disk tracer, because it quantifies the magnitude of the $K$-band excess above the photospheric level. Negative values of 
$\Delta(I_{\rm C}-K)$ are attributed by the authors to photometric variability or to errors either in photometry or in spectral 
types. The vertical line sets the limit of $\Delta(I_{\rm C}-K)$=0.3 proposed as the dividing line between disk-less and disked stars 
(\citealt{hillenbrandetal98}). It is clear the difference in the rotational behavior between stars with disk ($\Delta(I_{\rm C}-K)>0.3$) and 
those without disk ($\Delta(I_{\rm C}-K)<0.3$): while for disk-less stars the rotation rates expand over a large range possibly indicating
different time-scales for the unlocking, disked stars mostly present low rotation rates, apart from JW172. 
The presence of substantial numbers of slow-rotating stars with little or no excess could indicate they may have just recently 
cleared their disks and have not yet spun up in response to contraction on their way to the ZAMS. In order to 
assign a confidence level to our result, a one-side 2$\times$2 Fisher's exact test\footnote{We used the following web calculator 
(\citealt{Langsrudetal2007}): http://www.langsrud.com/fisher.htm.} was used (\citealt{Agresti1992}). We choose the division between 
disk-less and disked stars ($\Delta(I_{\rm C}-K)=0.3$) and the division at $v\sin i/R=1.5$ rad/day between slow and fast rotators. 
We find a {\it p-value} of 0.094 as chance that random data would yield this trend, indicating a probability of correlation 
of 90.6\%.

Concerning Orion, \cite{stassunetal99} have challenged the disk-locking scenario claiming no correlation 
between rotation and infrared properties related to circumstellar disks. The issue was re-examined by \cite{herbstetal02} 
using the $I_{\rm C}-K$ color excess derived by \cite{hillenbrandetal98} as a disk tracer and their own rotation measurements. They 
found that the mean value of $\Delta(I_{\rm C}-K)$ for the rapidly rotating sample (i.e., rotation period shorter than 3.14 
days) is of 0.17 mag, whereas for the slowly rotating sample (i.e., rotation period longer than 6.28 days) this value rises to 0.55 mag 
well above the limit of $\Delta(I_{\rm C}-K)$=0.3. The trend observed in Fig.~\ref{fig:vsini_disk1} is in agreement with the 
results reported by \cite{herbstetal01,herbstetal02}.

According to the disk-locking theory, a star is thought to be anchored or locked to its disk via its magnetic field that threads 
the circumstellar disk. Accretion of disk material occurs along the magnetic field lines. Thus, correlations between accretion indicators 
such as H$\alpha$ equivalent width or veiling index and rotation are expected (e.g., \citealt{hartiganetal95, hillenbrandetal98}). 
\cite{sicilia05} carried out a careful analyses of the H$\alpha$ profiles for a sample of 237 members of the ONC. They show that although 
the distribution of $(I_{\rm C}-K)$-excess for CTTS and WTTS overlap, a $(I_{\rm C}-K)$-excess of 0.5 can be taken as a 
trustful criteria to distinguish between CTTS and WTTS. As in \cite{herbstetal02}, \cite{sicilia05} also found evidence for a 
different behavior in rotation between the CTTS and WTTS in line with that predicted by the disk-locking scenario.

\begin{figure}	%[b!]
\begin{center}
 \begin{tabular}{c}
  \resizebox{\hsize}{!}{\includegraphics{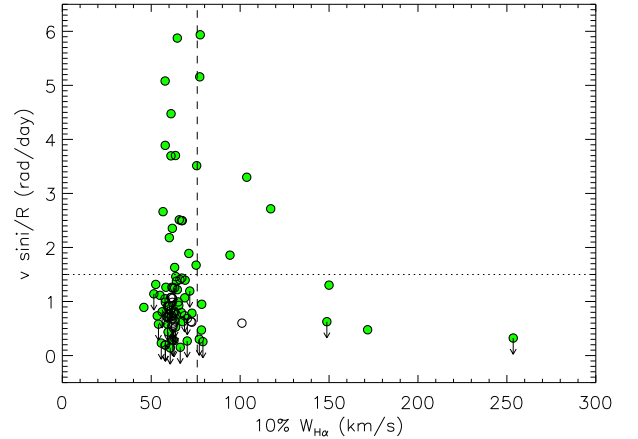}}  \\
 \end{tabular}
       \caption{Rotation against accretion tracers. Rotational velocity as a function of the full width of 
       the H$\alpha$ line at 10\% of its intensity. The empty circles represent the position of the most probable binaries. 
       The dashed line marks the maximum value of 10\% $W_{\rm H\alpha}$ measured in our sky spectra, while the 
       dotted line marks the division we considered for the analysis of the Fisher's test.}
       \label{fig:vsini_disk2}
 \end{center}
\end{figure}

As already pointed out in Sect.~\ref{sec:Halpha_line}, if the 10\% width criterion suggested by \cite{whitebasri03} is used to trace 
accretors, none of the stars in our sample is actually accreting. In Fig.~\ref{fig:vsini_disk2} we 
plot again the projected angular velocity as a function of the measured 10\% widths. The vertical line marks the 
maximum value of 10\% width measured in our sky spectra (at $\sim 76$ km s$^{-1}$). Therefore, stars with a 10\%-width 
less than this value are likely to be dominated by the sky emission. We promptly see that only 8 stars appear clearly to the 
right side of the vertical line. Two of them (JW937 and JW1036) show color excess $\Delta(I_{\rm C}-K)>0.3$. 
The other 4 slowly-rotating stars (JW52, JW296, JW559, and JW574) are sligthly on the right side of the vertical line, 
and with their values of $\sim 77-79$ km s$^{-1}$ seem to be not highly influenced by the sky fibers (Fig.~\ref{fig:eqw10}). 
Thus, even if all these objects are taken as {\it bona-fide} accretors, we can safely state that 
essentially only a fraction of about 14\% are accretors in our sample of $0.10-0.25 M_\odot$, which is lower than the 
range 61--88\% obtained by \cite{hillenbrandetal98} and 40--80\% obtained by \cite{sicilia05}. We do not 
believe this is due to an effect of sample selection, because our targets represent about 35\% of the \cite{hillenbrandetal98} 
sample with $0.10\le M \le 0.25M\odot$, available $\Delta(I_{\rm C}-K)$, and membership probability higher than 95\%. Moreover, 
our sample is well sampled in $\Delta(I_{\rm C}-K)$. In particular, considering the range $0.2<\Delta(I_{\rm C}-K)<0.5$ suitable 
as disk diagnostics, our sample represents about 25\% of the \cite{hillenbrandetal98} targets. Thus, in order to 
better understand our results, we considered the \cite{sicilia05} sample in the low-mass range $0.10-0.25M_\odot$, 
and we found a percentage of accretors of 13--26\%, which is more consistent with our findings. Thus, our results 
seem to be explained by assuming that low-mass stars are by percentage less active accretors than higher-mass 
stars. Relationships between accretion rates and mass of the central object have been investigated by several authors 
(see, e.g., the case of $\rho$ Ophiucus, where \cite{Nattaetal2006} find $\dot M_{\rm acc}\propto M_\star^{1.8}$ in 
$0.03-3M_\odot$ stars).

Contrary to the case of more massive stars where disk-locking is supported by rotation-color excess and rotation-accretion 
correlations (\citealt{herbstetal02, sicilia05}), the lack of strong concordance between Fig.~\ref{fig:vsini_disk1} 
and Fig.~\ref{fig:vsini_disk2} is intriguing. Dividing the data in Fig.~\ref{fig:vsini_disk2} in four boxes by 
$v\sin i/R=1.5$ rad/day and $10\% W_{\rm H\alpha}=76$ km s$^{-1}$, the one-sided 2$\times$2 Fisher's exact test leads to a {\it p-value} 
of 0.936, implying a probability of correlation of only 6.4\%.

\cite{herbstetal02} showed evidence in the ONC for a 
clear difference in the rotation behavior for stars below 0.25$M_\odot$ and those more massive than 0.25$M_\odot$ 
and this was also observed by \cite{Lammetal2005} for the older cluster NGC~2264. When the angular velocities of both 
groups are compared (Fig. 19 of \citealt{herbstetal02}), the distribution of angular velocity for the more massive stars is highly 
concentrated in one bin only at $\omega\sim1$ rad day$^{-1}$. This is in contrast to the angular velocity distribution for 
stars $M<0.25M_\odot$, which spread out essentially between $\omega\sim1-4$ rad day$^{-1}$. A possible interpretation given by 
\cite{herbstetal02} for this difference is that low-mass stars remained locked to their disks for a shorter period of 
time as compared to the more massive group. Thus, the distribution seen in ONC for the low-mass stars is already the result of 
the spin-up of a group of unlocked stars. Similar findings are reported by \cite{Lammetal2005} for NGC~2264.

\begin{figure*}
  \resizebox{\hsize}{!}{\includegraphics[bb=40 256 432 608,clip]{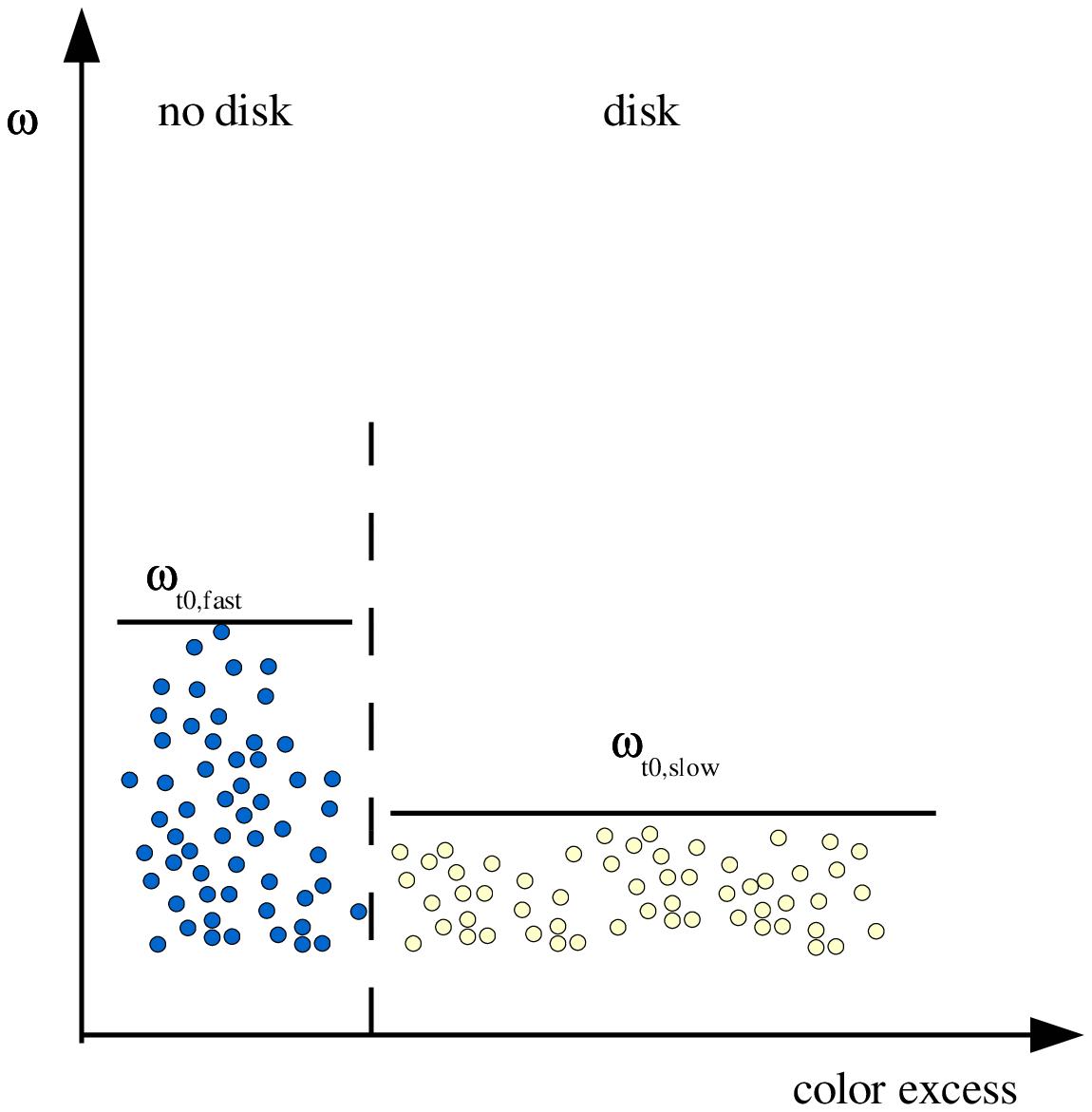}\includegraphics[bb=40 256 432 608,clip]{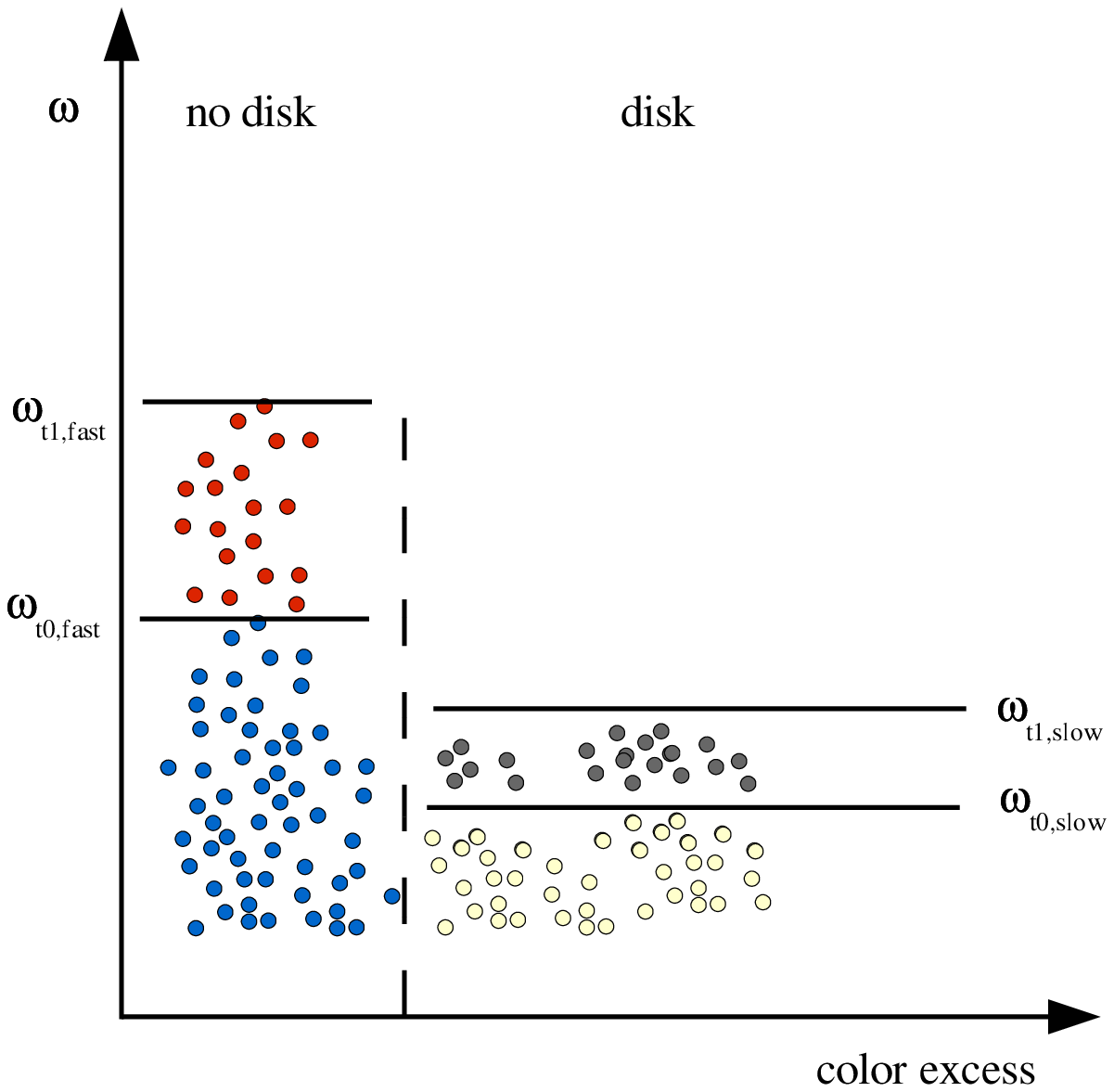}
  \includegraphics[bb=40 256 432 608,clip]{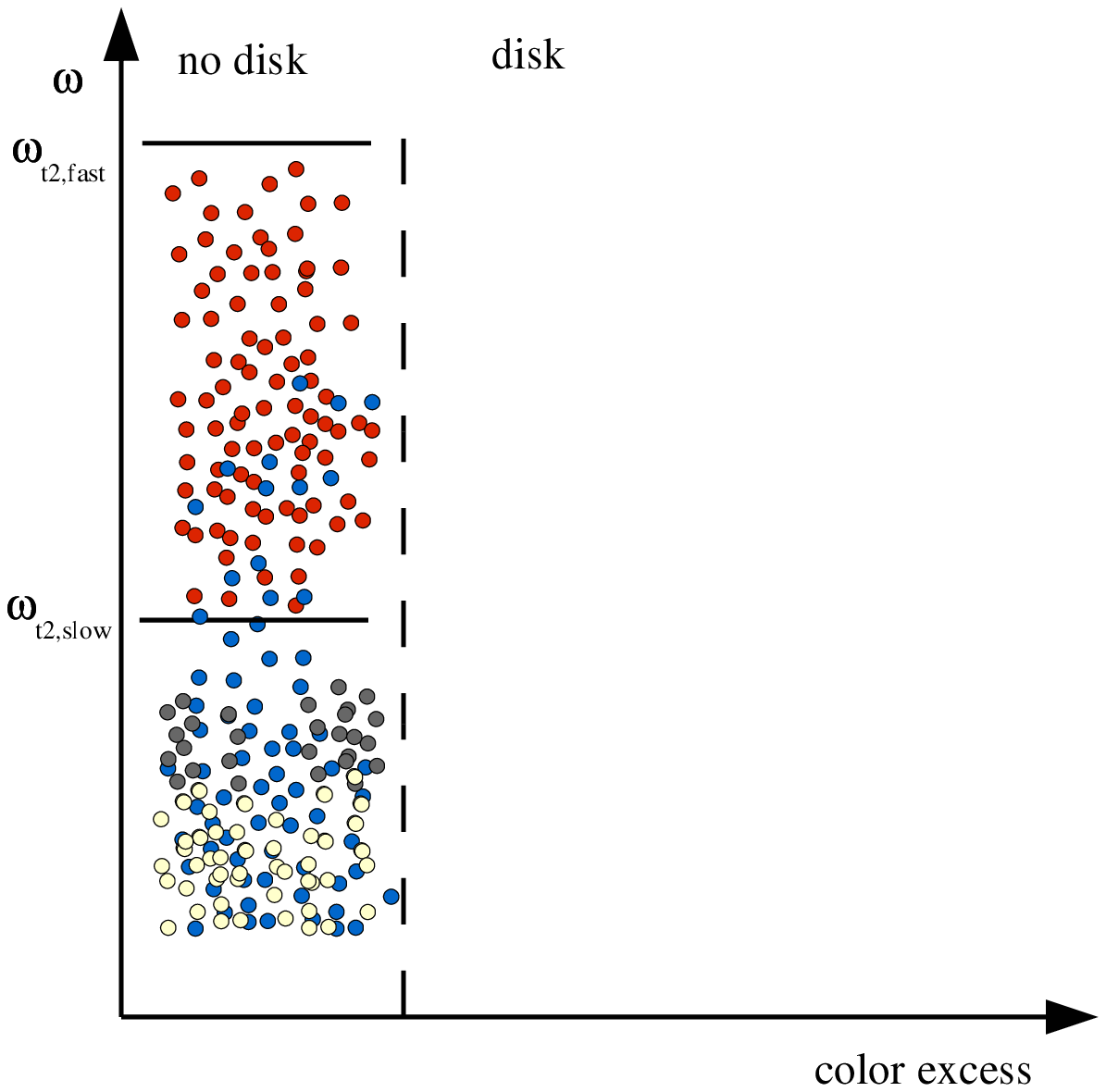}}
  %\resizebox{\hsize}{!}{\includegraphics[bb=40 256 432 608,clip]{./vsini_delta_ik_t1.eps}}
  %\resizebox{\hsize}{!}{\includegraphics[bb=40 256 432 608,clip]{./vsini_delta_ik_t2.eps}}
       \caption{Schematic view of the diagram angular rotational velocity versus color excess for three different evolutive times.}
       \label{fig:rot_evol}
\end{figure*}

Fig.~\ref{fig:vsini_disk2} suggests that indeed almost none of the stars observed here are locked to their disks. Are 
Fig.~\ref{fig:vsini_disk1} and Fig.~\ref{fig:vsini_disk2} in contradiction? In other words, can unlocked stars (i.e., 
non accreting stars) still hold a rotation-color-excess relation? In Fig.~\ref{fig:rot_evol} we schematically describe 
the evolution of a rotation ($\omega$) and color-excess (disk) diagram. Disk-locking operates for a fixed duration of 
time long enough to establish a rotation-color-excess relation. At a time $t_0$ all stars are unlocked 
(Fig.\ref{fig:rot_evol}a). As the contractions proceed all stars (slow and fast rotators) spin-up by a factor proportional 
to the value $(R_{t_0}/R_{t_1})^2$. Although color-excess diminishes due to disk dissipation, a rotation-color-excess 
relation is even stronger since the spin-up depends on $\omega_0$. Eventually disks disappear completely and we end up 
with a large spread in rotation as observed in the ZAMS clusters. Therefore, our Fig.~\ref{fig:vsini_disk1} and 
Fig.~\ref{fig:vsini_disk2} indicate that: {\it i) the low-mass stars in the ONC are not currently locked, but 
ii) the rotation-color-excess relation suggests that these stars were locked once.}

\begin{figure}	%[b!]
\begin{center}
 \begin{tabular}{c}
  \resizebox{\hsize}{!}{\includegraphics{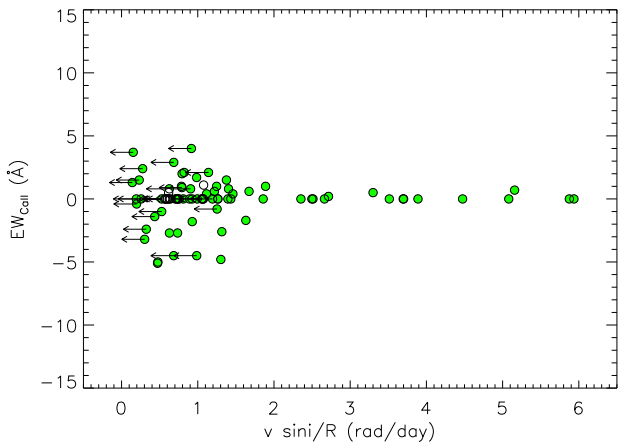}}  \\
 \end{tabular}
       \caption{\ion{Ca}{ii} $\lambda$8542 equivalent width (disk tracer) as a function of $v\sin i/R$. The possible 
       relation between disk frequency and rotation is not visible. The empty circles represent the position of the most probable binaries. }
       \label{fig:vsini_disk3}
 \end{center}
\end{figure}

From Fig.~\ref{fig:vsini_disk3}, the EW of $\lambda$8542 \ion{Ca}{ii} seems to have values indicating the presence 
of broad (also of the order of one hundred km s$^{-1}$) \ion{Ca}{ii} line emission or a partially filled-in line, 
phenomena associated with an accreting disk, for low $v\sin i/R$, while for higher rotational velocities there is 
almost no evidence of line emission or filling-in. \cite{hillenbrandetal98} conclude that stars with higher accretion rates should 
exhibit detectable \ion{Ca}{ii} emission above continuum levels, while stars that lack accretion disks should show net \ion{Ca}{ii} 
absorption with absorption equivalent widths $W_{\ion{Ca}{ii}}>1$. Possible explanations of the observed descrepancy 
between Fig.~\ref{fig:vsini_disk1} and Fig.~\ref{fig:vsini_disk3} could include the difficulty of measuring 
$W_{\ion{Ca}{ii}}$ for very late-type stars due to a low spectral resolution and to contrast effects. Moreover, as shown by 
\cite{batalhaetal96}, one expects a strong correlation between optical veiling (i.e., accretion) and the broad component 
of the \ion{Ca}{ii} line. In addition, at the spectral resolution used by \cite{hillenbrandetal98}, the broad component and the 
narrow component of this line always appear blended. Finally, small emission-line equivalent widths are more difficult to detect 
for low-mass stars because of the photospheric calcium line and the strong TiO bands. Thus, we think this diagnostic at present 
is not reliable and should be revisited using higher spectral resolution. Nevertheless, the fact that indeed most of the stars 
with broader \ion{Ca}{ii} line are those with $v \sin i < 9$ km s$^{-1}$ (our detection limit) prevent us in connecting 
this diagnostic to the disk-locking scenario. 

\begin{figure}	%[b!]
\begin{center}
 \begin{tabular}{c}
  \resizebox{\hsize}{!}{\includegraphics{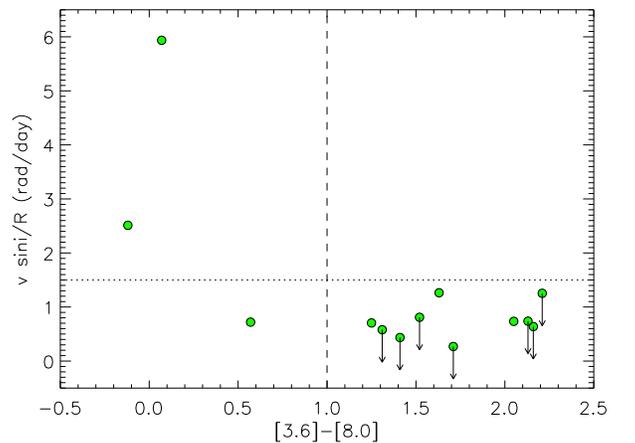}}  \\
 \end{tabular}
       \caption{Rotation against disk tracers. [3.8]$-$[8.0] as a function of $v\sin i/R$. A possible relation between 
       disk frequency and rotation is visible. The vertical line marks the boundary between disk and non-disk candidates, 
       while the orizontal line marks the division for the Fisher's test.}
       \label{fig:vsini_disk4}
 \end{center}
\end{figure}

Fig.~\ref{fig:vsini_disk4} shows the [3.8]$-$[8.0] color versus $v\sin i/R$ for the stars in common with \cite{rebulletal2006}. The number 
of stars is very limited (13), so it would seem conclusions could not be drown. We notice, however, that all the stars 
with strong color excess (i.e. above [3.8]$-$[8.0]=1) are slow rotators, and 7/10 have only detection limit $v\sin i$ measurements. 
Applying the one-sided Fisher's exact test, after the division of the plot at $v \sin i/R=1.5$ km s$^{-1}$ and 
[3.8]$-$[8.0]=1, we find a {\it p-value} of 0.038, which supports an association between the data at the $\sim$96\% confidence level 
(i.e. more than 2 sigma) even though the numbers of measurements is small. \cite{edwardsetal93} and 
\cite{rebulletal2006} found similar results and speculated that the long-period diskless objects have recently released 
their disks (within a few hundred thousand years). 

\subsubsection{Rotation and X-ray activity}

\begin{figure}[t]
  \resizebox{\hsize}{!}{\includegraphics{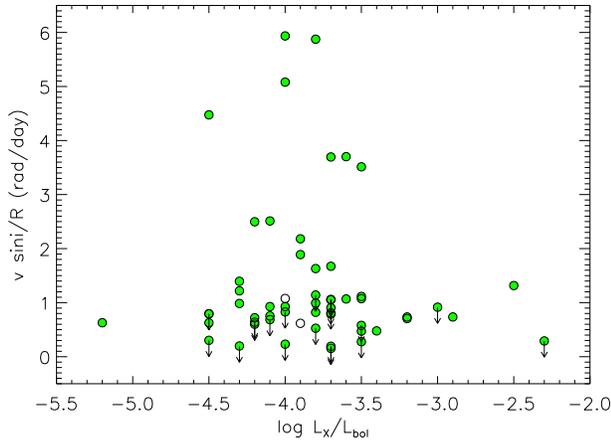}}
       \caption{Rotation versus activity for our stars. In spite of the scatter seen in the X-rays levels, 
        the data suggest that all stars in the sample are close to a saturated regime down to our detection limit in $v\sin i$. 
	The empty circles represent the position of the most probable binaries.}
       \label{fig:vsini_xrays}
\end{figure}

Based on X-ray measurements carried out by {\it Chandra} in the ONC, \cite{feigelsonetal03} have shown that the ONC 
stars have stronger X-ray emission than main-sequence stars with similar rotation periods and that the strong 
anti-correlation between X-rays and period observed for main-sequence stars is not seen for the ONC population. Instead, 
they claim that a marginal correlation (i.e., X-rays increasing with the rotation period) is seen in the data. In 
Fig.~\ref{fig:vsini_xrays} we show the logarithm of the ratio $L_{\rm X}/L_{\rm bol}$ of total band X-ray to bolometric 
luminosity as a function our $v\sin i$ divided by the theoretical radii for the stars in our sample in common to 
\cite{feigelsonetal03}. In spite of the scatter seen in the X-rays levels, the X-rays-rotation relation seems to be flat, 
suggesting that all stars in the sample are close to a ``saturated regime'' down to our detection limit in $v\sin i$. Recently, 
\cite{mohantybasri03} find that the chromospheric H$\alpha$ activity saturates at $\sim 3$ km s$^{-1}$ and at $\sim10$ km s$^{-1}$ 
for M4-M5- and for M5.5-M8.5-field dwarfs, respectively, while the X-ray activity saturates at $\sim3$ km s$^{-1}$ for early M 
stars. Considering their results and our $v\sin i$ detection limit, the unsaturated part of the possible activity-rotation 
relation is clearly beyond our sensitivity. 

For very low-mass stars (late M and L spectral types), the rotation-activity relationship is not really known, because 
their low rotation velocity is often below the detection limits (e.g., \citealt{stauffer97,mohantybasri03,messina07}).

\section{Summary and Conclusion}
\label{sec:summ_concl}
We have presented the results of a search for evidence of the disk-locking scenario as a mechanism for angular 
momentum depletion in a region centered on the Trapezium and extending beyond the ONC. We use our spectroscopic observations to 
derive radial velocities, and rotation rates, and to identify possible stars undergoing active disk accretion. In addition, 
we use IR photometric data from the literature to identify those stars possessing circumstellar disks. The results for our most 
probable single low-mass ($0.10<M<0.25M_\odot$) stars can be summarized as follows.
\begin{enumerate}
\item[-] We find a correlation between $v\sin i$ and $v_{\rm eq}$ confirming that the observed modulation is due to migrating 
inhomogeneities (\citealt{rhodeetal01}).
\item[-] We measure an average $\sin i$ lower than the mean value expected for a random distribution of stellar rotation axes, as already 
found by other authors (see, e.g., \citealt{rhodeetal01}).
\item[-] We do not find evidence for a population of fast rotators close to the break-up velocity. This supports the suggestions by 
\cite{hartmann02} that the timescale for angular momentum loss is comparable to the age of the ONC stars, so that disk braking may not 
be completely effective, at least among the lowest mass stars.
\item[-] From the $v\sin i-\Delta(I_{\rm C}-K)$ relation we find: 1) slow-rotators with disks, where the low $v\sin i$ is due 
to the coupling between star and disk; 2) slow-rotators without disks, as found by \cite{rebulletal2006} and corresponding 
to stars that are losing their disks, and hence continuing to rotate slowly for some time, but gradually spinning-up as they 
contract towards the ZAMS; 3) fast rotators without disks, where stars get free from their disks becoming fast rotators; 
4) almost no fast rotators with disks (with the exception of a single star).
\item[-] From the $v\sin i-10\%$-width relation, we do not find stars are actually accreting, but we argue that all non-accretors 
are either low or fast rotators, as found by \cite{jayawardhanaetal2006} in $\eta$ Cha and TW Hydrae ($\sim 6-8$ Myr).
\item[-] The \ion{Ca}{ii} line indicates possible phenomena associated with an accreting disk in stars around our $v\sin i$ 
limit, but we are very cautious about this diagnostic at present. A revision at higher resolution to arrive 
at a firmer conclusion.
\item[-] The few [3.8]$-$[8.0] colors available for our sample show that all stars with color excess are slow rotators.
\item[-] There is no clear relation between rotation and X-ray activity essentially due to the fact that all of our stars 
occupy ``saturated regime''.
\end{enumerate}
All these results are consistent with the following picture:
\begin{enumerate}
\item[-] very low-mass stars in our sample are not locked now, but they were locked in the past;
\item[-] the percentage of accretors seems to scale inversely to the mass of the stars.
\end{enumerate}

\begin{acknowledgements}
The authors are very grateful to the referee William Herbst for a careful reading of the paper and constructive 
suggestions. KB has been supported by the ESO DGDF 2008, and by the Italian {\em Ministero dell'Istruzione, Universit\`a e Ricerca} 
(MIUR). CHFM is very obliged to the {\sc flames} Science Verification team. This research has made use of the SIMBAD database, 
operated at CDS (Strasbourg, France), and the WEBDA database, operated at the Institute for Astronomy of the University 
of Vienna. This research made use of Montage, funded by the National Aeronautics and Space Administration's Earth Science 
Technology Office, Computational Technnologies Project, under Cooperative Agreement Number NCC5-626 between NASA and the 
California Institute of Technology. The code is maintained by the NASA/IPAC Infrared Science Archive.
\end{acknowledgements}

\bibliographystyle{aa}

\clearpage

\scriptsize
\begin{sidewaystable*}[f]
\caption{Our sample.}
\label{tab:sample}
\begin{center}
\begin{tabular}{rrrrrrcrrrrrrrrl}
\hline\hline             
JW	&  $I_{\rm C}$ & \tiny{$V-I_{\rm C}$} & \tiny{$\log T_{\rm eff}$}  & $R$ & $M$ & Spec. Type & \tiny{$\Delta (I_{\rm C}-K)$}  & $W_{\rm CaII}$  & \tiny{10\% $W_{\rm H\alpha}$} & $v\sin i$  
& $V_{\rm rad}$ & $\log\frac{L_{\rm X}}{L_{\rm bol}}$ & $P_{\rm rot}$ & \tiny{[3.6]$-$[8.0]}& Source \\
  	& (mag) &  &  (K) & ($R_{\odot}$) & ($M_{\odot}$) &    &  & (\AA)  & \tiny{(km s$^{-1}$)} & (km s$^{-1})$  
& (km s$^{-1}$) &  & (days) & & \\
\hline
  38 &  15.6 & 3.73 & 3.494 &  1.48 & 0.14 &	  M5	  & $-$0.35 & $-$1.0	&  62.2 &   $<$9.0	 &     24.80$\pm$0.28	&	 &	&    & 1;7,8\\
  43 &  15.8 & 3.73 & 3.483 &  1.27 & 0.11 &	M5.5	  & $-$0.33 &	 1.5	&  64.2 &  14.1$\pm$2.6  &     27.21$\pm$0.69	&	 &	&    & 1;7,8\\
  48 &  15.5 & 4.60 & 3.471 &  2.49 & 0.14 &	  M6	  & $-$0.69 &	 1.0	&  62.1 &  25.0$\pm$7.7  &     27.10$\pm$4.66	&	 &	&    & 1;7,8\\
50$^{*,\rm a}$& 11.7 & 1.75 & 3.602 &  3.46 & 0.34 &K1e,G-Ke,K7.3 &    1.37 & $-$14.6	&	&  20.9$\pm$1.2  &  $-$20.87$\pm$1.20	&	 &	&    & 1,2,3;7,8\\
     &       &      &	    &	    &	   &		  &	    &		&	&  23.8$\pm$1.4  &     66.38$\pm$1.40	&	 &	&    &  	 \\
     &       &      &	    &	    &	   &		  &	    &		&	&		 &22.75$\pm$1.80$^{\rm b}$  &	 &	&    &  	 \\
  52 &  14.6 & 3.64 & 3.494 &  2.25 & 0.15 &	  M5	  & $-$0.03 &	 0.0	&  79.1 &   $<$9.0	 &     24.43$\pm$0.54	&	 &	&    & 1;7,8	 \\
54$^{\rm a}$ &  15.3 & 3.58 & 3.494 &  1.57 & 0.14 &	  M5	  & $-$0.04 &	 0.0	&  62.2 &   $<$9.0	 &	4.80$\pm$0.25	&	 &	&    & 1;7,8\\
     &       &      &	    &	    &	   &		  &	    &		&	&  18.8$\pm$1.1  &     42.01$\pm$1.07	&	 &	&    &  	 \\
     &       &      &	    &	    &	   &		  &	    &		&	&		 &23.40$\pm$1.10$^{\rm b}$  &	 &	&    &  	 \\
  70 &  15.3 & 3.89 & 3.494 &  1.96 & 0.14 &	  M5	  & $-$0.35 &	 0.0	&  65.8 &  39.7$\pm$8.7  &     19.38$\pm$4.72	& $-$4.1 & 1.50 & $-$0.12 & 1;9,7,8,13	\\ 
99$^{*,\rm c}$&  13.1 & 2.51 & 3.562 &  2.92 & 0.24 &	  M1	  &    0.50 &	 1.4	&	&  16.9$\pm$0.9  &     66.34$\pm$0.93	&	 & 1.70 &    0.57 & 1;9,7,8,13\\
% 105 &  15.9 &	& 3.471 &  1.29 & 0.09 &      M6      & 	&    0.0    &  72.1 &  18.8$\pm$32.8 &  $-$34.80$\pm$71.64  & $-$4.1 &      &	 & 1;7,8    &	  No CCF\\ 
 110 &  16.6 &      & 3.483 &  0.78 & 0.09 &	M4-7	  &	    &	 0.0	&  61.9 &   $<$9.0	 &     28.01$\pm$0.96	& $-$3.7 &	&    & 1;7,8	\\ 
 118 &  14.8 & 3.09 & 3.483 &  1.79 & 0.13 &	M5.5	  & $-$0.32 &	 0.0	&  59.6 &  13.4$\pm$3.0  &     24.49$\pm$0.35	& $-$4.0 & 1.07 &    & 1;9,7,8  \\
 120 &  14.3 & 3.01 & 3.500 &  1.80 & 0.16 &	M4.5	  & $-$0.06 &	 1.0	&  71.2 &  27.4$\pm$5.3  &     22.51$\pm$1.25	& $-$3.9 &	&    & 1;11,7,8       \\ 
122$^{*}$ &  14.4 & 3.02 & 3.494 &  1.85 & 0.14 &	  M5	  & $-$0.17 &	 0.0	&  75.6 &  52.4$\pm$7.3  &     11.68$\pm$1.98	& $-$3.5 & 0.98 &    & 1;9,7,8  \\
 145 &  15.9 & 3.18 & 3.483 &  1.10 & 0.10 &	M5.3	  &    0.55 &		&  62.1 &   $<$9.0	 &     27.72$\pm$0.52	& $-$2.3 & 2.10 &    & 3;9,7,8  \\
 147 &  14.8 & 3.27 & 3.494 &  1.57 & 0.14 &	  M5	  &    0.06 &	 0.0	&  61.0 &  13.6$\pm$1.3  &     26.05$\pm$0.93	& $-$3.5 &	&    & 1;7,8	   \\
 155 &  14.9 & 3.26 & 3.500 &  1.52 & 0.15 &	M4.5	  &    0.15 &	 2.9	&  70.3 &   $<$9.0	 &     21.56$\pm$1.86	& $-$4.1 &	&    & 1;7,8	   \\
168$^{\rm d}$ &  15.0 & 3.16 & 3.483 &  1.64 & 0.12 &	M5.5	  & $-$0.06 &	 0.0	&  77.5 &  78.4$\pm$10.5 &	5.10$\pm$4.10	& $-$4.0 & 0.92 &    0.07 & 1;9,7,8,13 \\
 172 &  16.2 & 3.45 & 3.500 &  0.99 & 0.14 &	M4.5	  &    1.41 &	 0.0	&  63.6 &  29.5$\pm$2.7  &     26.71$\pm$4.63	& $-$3.6 & 1.43 &    & 1;9,7,8  \\
 179 &  16.6 &      & 3.483 &  0.77 & 0.09 &	M5.5	  &	    &	 0.0	&  71.7 &   $<$9.0	 &     25.43$\pm$1.01	&   &	   &	& 1;7,8    	 \\ 
 180 &  14.4 & 2.94 & 3.494 &  1.84 & 0.14 &	  M5	  &    0.05 &	 0.2	& 117.2 &  40.3$\pm$1.3  &     16.33$\pm$1.73	&   &	   &	& 1;7,8    \\ 
 183 &  14.5 & 3.45 & 3.494 &  2.05 & 0.15 &	  M5	  &    0.20 &	 0.0	&  67.0 &  41.2$\pm$2.6  &     22.14$\pm$2.17	& $-$4.2 & 1.51 &    & 1;9,7,8\\
 185 &  16.1 & 3.33 & 3.483 &  1.01 & 0.10 &	M5.5	  & $-$0.11 &		&  67.5 &  20.3$\pm$5.0  &     25.21$\pm$1.78	&   & 1.80 &	& 1;9,7,8  \\
 189 &  15.2 & 3.43 & 3.494 &  1.43 & 0.14 & M4.5IV-5V    &    0.40 &	 0.0	&  68.8 &  12.3$\pm$2.2  &     23.89$\pm$1.06	& $-$3.6 &	&    & 1;7,8 \\ 
 190 &  15.3 &      & 3.471 &  1.67 & 0.11 &	  M6	  &	    &	 2.0	&  67.0 &  10.7$\pm$2.1  &     26.71$\pm$0.65	& $-$4.5 &	&    & 1;7,8   \\ 
 228 &  16.7 & 3.68 & 3.494 &  0.89 & 0.12 &	 M5e	  &    0.18 & $-$0.8	&  63.2 &   $<$9.0	 &     22.92$\pm$0.80	&   & 3.22 &	2.21 & 1;9,7,8,13  \\
 230 &  14.6 &      & 3.471 &  2.36 & 0.14 &	  M6	  &	    &	 0.0	&  67.7 &  11.9$\pm$1.1  &     28.90$\pm$0.76	& $-$4.5 &	&    & 1;7,8	    \\ 
 231 &  15.6 &      & 3.471 &  1.49 & 0.10 &	  M6	  &	    &	 0.0	&  54.1 &   $<$9.0	 &     26.31$\pm$1.13	&   &	   &	& 1;7,8     \\ 
 234 &  15.9 & 2.88 & 3.500 &  0.86 & 0.13 &	M4.5	  &    0.41 &	 0.0	&  69.1 &   9.7$\pm$4.0  &     23.60$\pm$1.54	& $-$4.3 &	&    & 1;7,8 \\ 
 237 &  15.3 & 3.22 & 3.509 &  1.31 & 0.17 &	  M4	  &    0.75 &	 0.0	&  56.1 &   $<$9.0	 &     25.07$\pm$1.28	& $-$3.7 & 5.22 &    1.52 & 1;9,7,8,13 \\
239$^{\rm a}$&  12.7 & 2.25 & 3.555 &  2.73 & 0.24 & M1.5,M4:	  &    0.38 &	 1.5	&	&  19.2$\pm$1.1  &   $-$2.29$\pm$1.09	&   & 4.45 &	& 1,6;9,7,8 \\
     &       &      &	    &	    &	   &		  &	    &		&	&  16.1$\pm$0.9  &     70.06$\pm$0.88	&	 &	&    &  	 \\
     &       &      &	    &	    &	   &		  &	    &		&	&		 &33.89$\pm$1.40$^{\rm b}$  &	 &	&    &  	 \\
244$^{**}$&  15.5 & 2.91 & 3.494 &  1.15 & 0.13 &	  M5	  &    0.15 & $-$1.7	&  63.2 &  15.1$\pm$1.6  &     20.78$\pm$1.27	& $-$3.8 &	&    & 1;7,8\\ 
 257 &  14.5 & 2.64 & 3.500 &  1.67 & 0.15 &	M4.5	  &    0.15 &	 0.0	&  63.4 &   $<$9.0	 &     26.28$\pm$0.68	& $-$4.2 & 5.22 &    2.16 & 1;9,7,8,13 \\ 
276$^{\rm a}$&  14.0 & 3.04 & 3.500 &  2.15 & 0.15 &  M4-4.5	  &    0.10 &	 1.1	&  61.5 &  18.6$\pm$1.1  &	4.41$\pm$1.05	& $-$4.0 &	&    & 1;7,8	\\	 
     &       &      &	    &	    &	   &		  &	    &		&	&  10.7$\pm$0.5  &     42.20$\pm$0.50	&	 &	&    &  	 \\
     &       &      &	    &	    &	   &		  &	    &		&	&		 &23.30$\pm$1.16$^{\rm b}$   &	 &	&    &  	 \\
% 282 &  16.3 &	& 3.500 &  0.75 & 0.12 &    M4.5      & 	&    0.0    &  63.8 & 9999	  ???		&	???	       & $-$2.9 & 0.78 &    & 1;9,7,8	& \\
 288 &  15.4 & 2.73 & 3.494 &  1.17 & 0.13 &	  M5	  &    0.26 &	 0.9	&  63.4 &   $<$9.0	 &     26.71$\pm$0.58	& $-$4.5 &	&    & 1;7,8	 \\    
 293 &  14.5 & 3.34 & 3.494 &  1.87 & 0.14 &	  M5	  &    0.13 &	 0.5	& 103.7 &  49.7$\pm$9.4  &     20.71$\pm$2.14	&   & 0.98 &	& 1;9,7,8   \\
 294 &  15.4 & 3.50 & 3.500 &  1.45 & 0.15 &	M4.5	  &    0.05 &	 0.0	&  67.2 &  16.7$\pm$2.5  &     25.74$\pm$1.05	&   & 2.57 &	& 1;9,7,8   \\
 296 &  14.3 & 3.31 & 3.500 &  2.10 & 0.16 &   M4.5e	  &    0.67 & $-$3.2	&  77.0 &   $<$9.0	 &     23.52$\pm$0.45	& $-$4.5 &	&    & 1;7,8\\    
297$^{\rm e}$ &  16.7 &      & 3.483 &  0.75 & 0.09 &	M5.5	  &	    &	 0.0	&  56.6 &  16.1$\pm$2.8  &     26.25$\pm$4.92	&   &	   &	& 1;7,8  \\ 
 298 &  16.1 &      & 3.471 &  1.20 & 0.09 &	  M6	  &	    &		&  60.8 &   $>$9.0	 &     31.33$\pm$7.60	& $-$4.0 &	&    & 1;7        \\ 
 304 &  15.1 & 3.26 & 3.494 &  1.39 & 0.13 &	  M5	  &    0.35 &	 0.6	&  75.2 &  18.7$\pm$1.6  &     22.57$\pm$1.30	& $-$3.7 & 3.03 &    & 1;9,7,8 \\
 318 &  15.9 & 3.32 & 3.500 &  0.99 & 0.14 &	M4.5	  &    0.42 &	 0.8	&  66.1 &  11.2$\pm$1.4  &     31.17$\pm$0.68	&   & 3.40 &	& 1;9,7,8   \\ 
322$^{**}$&  15.0 &      & 3.500 &  1.33 & 0.15 &	M4.5	  &	    & $-$1.8	&  65.5 &   9.9$\pm$3.7  &     29.53$\pm$1.06	& $-$4.1 &	&    & 1;7,8\\ 
 333 &  14.7 & 1.92 & 3.483 &  1.89 & 0.13 &	M5-6	  &    0.54 &	 3.7	&  66.3 &   $<$9.0	 &     27.10$\pm$0.61	& $-$3.7 &	&    & 1;7,8	     \\ 
\hline										       
\end{tabular}									       
\end{center}
\end{sidewaystable*}

\begin{sidewaystable*}[f]
{\bf Table 2.} Continued.\\
\begin{center}
\begin{tabular}{rrrrrrcrrrrrrrrl}
\hline\hline             
JW	& $I_{\rm C}$ & \tiny{$V-I_{\rm C}$} & \tiny{$\log T_{\rm eff}$}  & $R$ & $M$ & Spec. Type & \tiny{$\Delta (I_{\rm C}-K)$}  & $W_{\rm CaII}$  & \tiny{10\% $W_{\rm H\alpha}$}& $v\sin i$  
& $V_{\rm rad}$ & $\log\frac{L_{\rm X}}{L_{\rm bol}}$ & $P_{\rm rot}$ & \tiny{[3.6]$-$[8.0]} & Source\\
  	& (mag) &  &  (K) & ($R_{\odot}$) & ($M_{\odot}$) &    &  & (\AA)  & \tiny{(km s$^{-1}$)} & (km s$^{-1})$  
& (km s$^{-1}$) &  & (days) & & \\
\hline
 353 &  14.9 & 3.73 & 3.494 &  2.06 & 0.15 &	  M5	  &    0.13 &	 1.3	&  60.7 &    $<$9.0	  & 24.63$\pm$0.84   &        &      &         & 1;7,8     \\ 
 357 &  14.3 & 3.02 & 3.500 &  1.83 & 0.16 &	M4.5	  & $-$0.00 &	 0.0	&  65.1 &   14.6$\pm$1.5  & 26.69$\pm$0.66   & $-$3.8 &      &         & 1;7,8       \\  
 366 &  15.0 & 3.98 & 3.497 &  2.43 & 0.15 & M4.5-5e	  &    0.05 & $-$2.7	&  72.4 &   12.3$\pm$3.8  & 22.89$\pm$1.18   & $-$5.2 &      &         & 1;7,8,10  	 \\ 
 379 &  15.2 & 3.87 & 3.483 &  1.84 & 0.13 &	M5.4	  &    0.07 &		&  70.2 &    $<$9.0	  & 24.03$\pm$0.51   &        & 1.30 &    1.71 & 3;9,7,8,13 \\ 
 389 &  15.8 & 3.64 & 3.488 &  1.21 & 0.12 &  M4.5-6	  &    1.44 &	 4.0	&  60.3 &    $<$9.0	  & 22.06$\pm$2.05   & $-$3.0 & 5.54 &         & 1;9,7,8,10  \\ 
 392 &  15.2 &      & 3.465 &  1.93 & 0.11 &  M6-6.5	  &	    &	 0.0	&  63.7 &    $<$9.0	  & 23.55$\pm$0.63   & $-$3.8 &      &         & 1;7,8      \\ 
 402 &  15.3 & 3.73 & 3.483 &  1.62 & 0.12 &	M5.5	  &    0.72 & $-$1.4	&  59.5 &    $<$9.0	  & 25.74$\pm$1.11   &        & 5.00 &    1.41 & 1;9,7,8,13   \\ 
 404 &  14.7 & 3.38 & 3.494 &  1.81 & 0.14 & M4.5-5.5	  &	    &		&  72.9 &   11.4$\pm$1.8  & 25.99$\pm$1.44   &        &      &         & 1;7,10       \\ 
446$^{\rm f}$&  16.2 & 2.98 & 3.500 &  0.77 & 0.13 &	M4.5	  &    0.04 &	 0.0	&  64.6 &   36.4$\pm$10.7 & 20.92$\pm$3.73   & $-$3.8 &      &         & 1;7,8     \\ 
 469 &  14.5 & 2.88 & 3.494 &  1.79 & 0.14 & M4.5-5.5	  & $-$0.03 &	 0.0	&  64.2 &   10.9$\pm$1.6  & 29.69$\pm$0.85   & $-$4.1 &      &         & 1;7,8        \\ 
 486 &  15.4 & 3.82 & 3.500 &  1.81 & 0.16 &	M4.5	  & $-$0.07 &	 0.0	&  62.4 &    $<$9.0	  & 27.10$\pm$0.48   & $-$4.2 &      &         & 1;7,8       \\ 
% 490 &  15.0 & 4.10 & 3.494 &  2.55 & 0.15 &      M5      &    1.08 & $-$1.2     & 59.5   	   & ???	  &	       ???	 & $-$4.4 &	 &	   & 1;7,8,10	& $^{**}$\\ 
500$^{\rm a}$&  14.8 & 3.71 & 3.483 &  1.97 & 0.13 &	M5.5	  &    0.19 &	 0.5	&  72.9 &    9.8$\pm$0.4  & 17.40$\pm$0.44   & $-$3.9 &      &         & 1;7,8,10\\
     &       &      &	    &	    &	   &		  &	    &		&	&    $<$9.0	  & 43.63$\pm$0.14   &        &      &         &	   \\
     &       &      &	    &	    &	   &		  &	    &		&	&		  & 30.52$\pm$0.46$^{\rm b}$  &   &      &         &	   \\
 510 &  15.3 & 3.93 & 3.494 &  2.01 & 0.14 &	  M5	  & $-$0.10 & $-$0.4	&  57.8 &    3.2$\pm$3.7  & 29.03$\pm$1.68   & $-$4.3 &      &         & 1;7,8,10  	    \\ 
 517 &  15.0 & 3.22 & 3.500 &  1.40 & 0.15 &	M4.5	  & $-$0.08 &	 0.0	&  61.1 &   50.5$\pm$9.6  & 21.98$\pm$3.92   & $-$4.5 & 0.85 &         & 1;11,7,8  	     \\ 
 523 &  15.8 & 3.75 & 3.500 &  1.43 & 0.15 &	M4.5	  &    0.04 &	 1.7	&  61.9 &   11.3$\pm$1.3  & 23.23$\pm$1.00   & $-$4.3 &      &         & 1;7,8        \\ 
 530 &  15.3 & 3.44 & 3.494 &  1.40 & 0.13 &	  M5	  & $-$0.22 &	 0.7	&  77.3 &   58.2$\pm$0.8  & 20.69$\pm$2.30   &        &      &         & 1;7,8     	    \\ 
 545 &  15.2 & 3.42 & 3.494 &  1.41 & 0.13 &	  M5	  & $-$0.09 &	 0.0	&  94.3 &   21.1$\pm$2.0  & 27.07$\pm$1.61   &        & 1.71 &         & 1;11,7,8  	    \\ 
 559 &  14.7 & 4.17 & 3.483 &  2.91 & 0.16 &   M5.5e	  & $-$0.31 & $-$5.1	&  78.2 &   11.1$\pm$2.1  & 25.19$\pm$1.30   & $-$3.5 & 2.27 &         & 1;12,7,8,10         \\ 
 573 &  15.9 & 2.83 & 3.500 &  0.89 & 0.13 &	M4.5	  &    1.41 &	 0.0	&  57.8 &    $<$9.0	  & 23.79$\pm$0.94   & $-$3.7 &      &         & 1;7,8     	     \\ 
 574 &  14.8 & 2.88 & 3.500 &  1.45 & 0.15 &	M4.5	  &    3.17 &		&  78.4 &   11.9$\pm$1.3  & 24.08$\pm$1.51   &        &      &         & 1;7,8     	     \\ 
 577 &  15.3 & 3.58 & 3.483 &  1.43 & 0.11 &	M5.5	  &    0.17 &	 1.0	&  62.4 &    9.1$\pm$1.3  & 23.33$\pm$0.56   & $-$3.7 &      &         & 1;7,8        \\ 
 615 &  16.3 &      & 3.471 &  1.05 & 0.08 &	  M6	  &	    & $-$4.5	&  57.9 &    $<$9.0	  & 23.26$\pm$1.33   &        &      &         & 1;7,8        \\ 
 635 &  15.9 & 4.10 & 3.500 &  1.80 & 0.16 &	M4.5	  &    2.16 &	 0.0	&  58.3 &   18.3$\pm$0.9  & 27.82$\pm$2.06   &        & 4.12 &    1.63 & 1;9,7,8,13   \\ 
 647 &  13.7 & 2.31 & 3.494 &  2.66 & 0.15 &	 M5e	  &	    & $-$5.0	& 171.7 &   10.2$\pm$1.1  & 24.79$\pm$1.17   & $-$3.4 &      &         & 1;7,8     	   \\ 
 653 &  15.5 & 3.32 & 3.465 &  1.70 & 0.11 & M6-M6.5	  & $-$0.12 &	 0.0	&  61.6 &   17.3$\pm$1.7  & 24.11$\pm$1.58   &        &      &         & 1;7,8       \\ 
 663 &  14.6 & 2.36 & 3.500 &  1.63 & 0.15 &	M4.5	  &    1.02 &	 2.4	&  62.8 &    $<$9.0	  & 27.82$\pm$0.50   & $-$3.5 & 1.31 &         & 1;9,7,8      \\ 
669$^{*, \rm d}$&  10.8 & 1.67 & 3.708 &  4.66 & 2.13 &G8,$<$K5,K3,K1-2&  0.14 &	 1.8	&	&   65.0$\pm$6.0  & 24.60$\pm$2.50   &        & 1.81 &        & 1,2,4,5;12,7,8,10\\
689$^{\rm c}$ &  16.9 &      & 3.483 &  0.69 & 0.08 &	M5.5	  &	    &	 0.0	&  57.9 &   21.6$\pm$1.3  & 21.84$\pm$1.26   &        &      &         & 1;7,8     \\
 715 &  15.3 & 3.46 & 3.471 &  1.70 & 0.11 &   M6,M5	  & $-$0.08 &	 0.0	&  59.2 &    $<$9.0	  & 26.03$\pm$0.99   & $-$3.5 &      &    1.31 & 1,6;7,8,13  \\ 
 716 &  14.4 & 2.88 & 3.500 &  1.73 & 0.15 &   M4.5e	  &    1.15 & $-$2.7	&  61.6 &   10.2$\pm$3.0  & 25.14$\pm$1.21   & $-$2.9 & 3.95 &    2.05 & 1;9,7,8,13   \\  
 755 &  15.1 & 3.79 & 3.460 &  2.18 & 0.11 &	M6.5	  & $-$0.09 &	 0.0	&  59.0 &   12.7$\pm$4.3  & 26.45$\pm$0.92   & $-$4.2 &      &         & 1;7,8      \\
 763 &  16.4 & 2.54 & 3.494 &  0.73 & 0.11 &	  M5	  &    0.97 &	 0.8	&  59.1 &    $<$9.0	  & 23.72$\pm$0.72   & $-$3.7 &      &         & 1;7,8      \\
 807 &  15.3 &      & 3.471 &  1.68 & 0.11 &	  M6	  &	    &	 2.1	&  63.1 &   11.1$\pm$2.1  & 24.84$\pm$0.64   & $-$3.8 &      &         & 1;7,8     \\
 817 &  13.8 & 3.04 & 3.500 &  2.37 & 0.15 &	M4.5	  &    0.22 &		&  60.2 &   41.6$\pm$3.8  & 24.50$\pm$6.11   & $-$3.9 & 2.23 &         & 3;9,7,8     \\
 823 &  13.9 & 3.28 & 3.494 &  2.40 & 0.15 &	  M5	  &    0.25 &	 0.0	&  58.0 &    $<$9.0	  & 25.58$\pm$0.53   & $-$3.7 & 9.00 &         & 1;9,7,8     \\
840$^{\rm a}$&  14.3 & 2.98 & 3.494 &  1.97 & 0.14 &	  M5	  &    0.17 &	 0.0	& 101.0 &    9.5$\pm$0.4  &$-$14.25$\pm$0.42 &        &      &         & 1;7,8  \\
     &       &      &	    &	    &	   &		  &	    &		&	&    5.3$\pm$0.1  & 61.74$\pm$0.12   &        &      &         &	   \\
     &       &      &	    &	    &	   &		  &	    &		&	&		  & 23.75$\pm$0.44$^{\rm b}$  &   &      &         &	    \\
 846 &  16.1 & 4.13 & 3.500 &  1.67 & 0.15 &	M4.5	  &    0.82 &	 0.0	&  45.9 &   12.0$\pm$2.9  & 26.13$\pm$2.20   &        & 5.46 &         & 1;9,7,8   	   \\
 859 &  15.0 &      & 3.494 &  1.44 & 0.14 &	  M5	  &	    & $-$2.6	&  52.5 &   15.3$\pm$2.4  & 25.39$\pm$1.05   & $-$2.5 & 3.60 &         & 1;9,7,8   	   \\
 862 &  15.5 & 3.80 & 3.471 &  1.57 & 0.11 &	 M6e	  &    0.24 & $-$4.5	&  62.1 &    $<$9.0	  & 27.33$\pm$0.67   &        & 3.49 &         & 1;9,7,8      \\
 865 &  14.3 & 3.60 & 3.494 &  2.45 & 0.15 &	  M5	  &    0.01 &	 1.5	&  55.7 &    $<$9.0	  & 23.81$\pm$0.43   & $-$4.0 &      &         & 1;7,8     	   \\
 872 &  15.2 & 3.24 & 3.494 &  1.32 & 0.13 &	  M5	  &    0.60 &	 0.4	&  54.9 &   11.9$\pm$1.6  & 26.42$\pm$2.50   & $-$3.5 & 4.19 &         & 1;9,7,8     \\
 878 &  14.4 & 3.74 & 3.471 &  2.51 & 0.14 &	  M6	  & $-$0.06 &	 0.0	&  59.0 &   14.3$\pm$1.7  & 27.03$\pm$1.14   & $-$3.2 & 5.54 &    1.25 & 1;9,7,8,13   \\
% 883 &  13.26& 2.62 &	&	&      &	      & 	&           &       &   36.1$\pm$36.1 &	67.74$\pm$67.74  &	  & 0.85 &	   & 2;7	& $^*$, No CCF\\
 898 &  14.1 & 3.30 & 3.494 &  2.22 & 0.15 &      M5      &    0.07 &    0.0    &  53.4 &   13.1$\pm$0.6  &	29.12$\pm$0.63   & $-$3.2 &	 &	   & 1;7,8   \\ 
 908 &  14.4 & 3.05 & 3.500 &  1.78 & 0.15 &    M4.5      &    0.02 &    0.8    & 148.8 &    $<$9.0	      &	28.25$\pm$0.58   &	  &	 &	   & 1;7,8    \\ 
\hline										     
\end{tabular}									     
\end{center}									     
\end{sidewaystable*}

\begin{sidewaystable*}[f]
{\bf Table 2.} Continued.\\
\begin{center}
\begin{tabular}{rrrrrrcrrrrrrrrrl}
\hline\hline             
JW	&  $I_{\rm C}$ & \tiny{$V-I_{\rm C}$} & \tiny{$\log T_{\rm eff}$}  & $R$ & $M$ & Spec. Type & \tiny{$\Delta (I_{\rm C}-K)$} & $W_{\rm CaII}$ &\tiny{10\% $W_{\rm H\alpha}$} & $v\sin i$ & $V_{\rm rad}$ & $\log\frac{L_{\rm X}}{L_{\rm bol}}$ & $P_{\rm rot}$ & \tiny{[3.6]$-$[8.0]} & Source \\
  	&  (mag) &  &  (K) & ($R_{\odot}$) & ($M_{\odot}$) &    &  & (\AA)  & \tiny{(km s$^{-1}$)} & (km s$^{-1})$  
& (km s$^{-1}$) &  & (days) & &  \\
\hline
 929 &  15.9 & 3.82 & 3.483 &  1.32 & 0.11 &	M5.5	  & $-$0.37 &	 0.0	& 61.1 &    39.3$\pm$5.9  &	25.47$\pm$3.24      & $-$3.7 & 0.81 &	      & 1;9,7,8 \\ 
 937 &  15.6 & 3.68 & 3.471 &  1.47 & 0.10 &	  M6	  &    0.89 & $-$4.8	&150.0 &    15.4$\pm$1.8  &	23.13$\pm$1.63      &	     &      &	      & 1;7,8	 \\ 
 939 &  16.4 & 2.56 & 3.500 &  0.72 & 0.12 &	M4.5	  &    0.84 &	 2.1	& 51.5 &     $<$9.0	  &	27.28$\pm$0.87      & $-$3.8 &      &	      & 1;7,8	 \\ 
 961$^{*, \rm c}$ &  12.3 & 1.71 & 3.612 &  2.47 & 0.45 &	K6-7	  &    0.18 &	 1.9	&      &    67.0$\pm$6.0  &	21.25$\pm$2.00      &	     & 1.43 &	      & 1;9,7,8 \\
 964$^{\rm c}$ &  14.7 & 3.59 & 3.483 &  1.95 & 0.13 &	M5.5	  &    0.34 &	 0.0	& 57.8 &    79.8$\pm$5.3  &	 9.54$\pm$5.34      & $-$4.0 &      &	      & 1;7,8	\\ 
 990 &  14.1 & 2.88 & 3.500 &  2.07 & 0.16 &	M4.5	  & $-$0.14 &	 0.6	& 64.7 &    20.3$\pm$0.7  &	27.67$\pm$0.44      & $-$4.3 & 2.16 &	      & 1;9,7,8 \\ 
1025 &  15.9 & 4.01 & 3.483 &  1.46 & 0.11 &	M5.5	  &    0.24 &	 0.0	& 68.7 &     $<$9.0	  &	27.18$\pm$1.11      &	     & 4.52 &	 2.13 & 1;9,7,8,13  \\ 
1032$^{\rm c}$ &  16.2 & 3.90 & 3.483 &  1.21 & 0.11 &	M5.5	  & $-$0.45 &	 0.0	& 61.9 &    22.9$\pm$1.4  &	31.19$\pm$1.35      &	     &      &	      & 1;7,8\\ 
1034 &  15.0 & 3.11 & 3.494 &  1.44 & 0.14 &	  M5	  & $-$0.15 &	 0.4	& 63.8 &    16.9$\pm$0.8  &	19.92$\pm$0.69      &	     &      &	      & 1;7,8	\\  
1036 &  15.8 & 3.74 & 3.471 &  1.36 & 0.09 &	  M6	  &    0.31 & $-$2.4	&253.5 &     $<$9.0	  &	28.32$\pm$0.83      &	     &      &	      & 1;7,8   \\  
\hline
\end{tabular}
\end{center}

{\sc Notes - }\\
Spectral type sources: 1=\cite{hillenbrand97}; 2=\cite{parenago1954}; 3=Prosser \& Stauffer, unpublished; 4=Samuel (1993), unpublished PhD thesis; 5=\cite{cohenkuhi79}; 
6=\cite{edwardsetal93}.\\
Astrophysical parameter sources: 7=\cite{hillenbrand97}; 8=\cite{hillenbrandetal98}; 9=\cite{herbstetal02}; 10=\cite{prosseretal1994}; 11=\cite{stassunetal99}; 
12=\cite{herbstetal00}; 13=\cite{rebulletal2006}.\\
$^*$: classified as binary by \cite{rhodeetal01}; $^{**}$: $I_{\rm C}-K$ excess contaminated by the Pa 8548 line (\citealt{hillenbrandetal98}).\\
This work: $^{\rm a}$: classified as binary; $^{\rm b}$: barycenter velocity of the binary system assuming equal mass components; $^{\rm c}$: only one measure; $^{\rm d}$: bad CCF; 
$^{\rm e}$: uncertain CCF; $^{\rm f}$: uncertain $v\sin i$.\\
\end{sidewaystable*}

\normalsize

\end{document}